\documentclass[11pt,a4paper]{article}
\usepackage{longtable}
\usepackage{amsmath}
\usepackage{amssymb}
\usepackage{graphicx}
\usepackage{bm}
\usepackage{footmisc}
\usepackage[section]{placeins}

\newcommand*{\bea}{\begin{eqnarray}}
\newcommand*{\eea}{\end{eqnarray}}
\newcommand*{\be}{\begin{equation}}
\newcommand*{\ee}{\end{equation}}

\newcommand{\bma}{\begin{pmatrix}}
\newcommand{\ema}{\end{pmatrix}}

\usepackage[square,comma,sort&compress,numbers]{natbib}
\title{Studies of Wick Cutkosky model in light of its classical ground state using Dyson Schwinger equations}
\author{Tajdar Mufti \footnote{tajdar.mufti@gmail.com, tajdar.mufti@sse.habib.edu.pk} \\ Habib University, Block 18, Gulistan-e-Jauhar \\ University Avenue, Off Shahrah-e-Faisal, Karachi 75290, Pakistan}

\date{}
\begin{document}
\maketitle
\begin{abstract}
Implications of additional fundamental scalar fields are regarded as among the most important avenues to explore after the experimental discovery of the standard model Higgs, particularly when there already exist cogent arguments in favor of their existence. A peculiar observation in Higgs-scalar singlet system is tendency of scalar singlet field to have negative squared physical masses which may be a sign of either a tachyon field or symmetry breaking. Assuming that this feature is due to the presence of a phenomenon similar to the conventionally understood Higgs mechanism, Wick Cutkosky model is studied in the parameter space suggested by the classical ground state of the system at Higgs mass $125.09$ GeV with positive values for both squared bare masses. The results are found to have strong negative contributions to squared scalar masses. For higher cutoff values, the renormalized squared scalar masses and squared bare couplings are found to qualitatively favor a relationship similar to the one in the classical ground state of the system, upto an additive constant. Higgs propagators remain almost unaffected as observed in a previously explored region of parameter space. However, scalar singlet propagators are found to have relatively different qualitative features in comparison to the previous study of the model. Vertices are found to have qualitatively similar features. No sign of triviality is found.
\end{abstract}
\section{Introduction}
Discovery of Higgs \cite{pdg,Maas:2017wzi,Carena:2002es} at LHC \cite{Aad:2012tfa,Chatrchyan:2012xdj} has certainly validated the standard model (SM) as the most reliable low energy theory, currently available, for phenomenology of particles as seen in accelerator experiments. Despite that there are certain expectations from Higgs in other quantum field theories, Supersymmetry (QFTs) \cite{Kane:2018oax,Haber:1989xc,Gunion:2002zf,Martin:1997ns} as an example, odds are currently in favor of the SM Higgs boson, particularly in absence of any experimental discovery at LHC \cite{Aaboud:2018doq,Aaboud:2017leg} related to physics beyond SM. However, as Higgs' discovery raises hopes for more scalar particles to be annexed in the SM or other QFTs, it also necessitates a number of questions to be addressed. Understanding possible extensions of the conventionally understood spontaneous symmetry breaking mechanism (SSBM) \cite{Englert:1964et,Higgs:1964pj,Higgs:1966ev,Kaku:1993ym,Peskin:1995ev} is among them. A QFT of scalar fields offer highly suitable playground for these explorations.
\par
There are a number of rationales to assume high prospects of other scalar fields in nature. In recent years, the idea of Higgs providing slow rollover during cosmic inflation \cite{Mukhanov:2005sc} caught considerable attention \cite{Guth:1980zm,Ferreira:2017ynu,Hakim:1984oya}. However, the quartic Higgs coupling is found to be around 0.6 \cite{Ferreira:2017ynu}, which is different from the expected value for a successful inflation by orders of magnitudes. It indicates that at least one more scalar should exist in nature to fulfill the role of inflaton. However, it is important to note that relevance of further scalars to inflationary scenarios is a decades old idea \cite{Linde:1993cn} in which both Higgs and another scalar field are considered for the so-called hybrid inflationary scenario. Furthermore, scalar fields are also regarded as contributors in the physics related to dark matter \cite{Lee:2017qve,Athron:2017kgt} beside supersymmetric particles \cite{Munoz:2017ezd,Takayama:2000uz}.
\par
One of the triumphs for (minimal) supersymmetry \cite{Martin:1997ns} is expecting the Higgs boson with mass less than 135 GeV \cite{Haber:1990aw,Ellis:1990nz,Okada:1990vk,Carena:2002es,Ellis:1992cp}, despite that supersymmetry is yet to be experimentally discovered \cite{Aaboud:2018doq,Aaboud:2017leg}. There is known a number of possibilities in supersymmetry, for instance \cite{Ellis:2000ig}, suggesting a number of scalars providing various extensions to the already known SM \cite{Ellis:2007wa}. Furthermore, supersymmetry has also been studied in the context of cosmological scenarios \cite{Jungman:1995df}.
\par
Despite simplicity of their structure, implications of scalar fields in QFTs are found highly non-trivial. Recently, in a variant \cite{Mufti:2018xqq} of Wick Cutkosky's model \cite{Darewych:1998mb,Sauli:2002qa,Efimov:2003hs,Nugaev:2016uqd,Darewych:2009wk} it was observed that in some regions of parameter space scalar singlet propagators have negative self energy terms which raised suspicion that in an extreme case the renormalized squared scalar mass may even flip the sign in some region in parameter space as in the well understood SSBM \cite{Kaku:1993ym,Peskin:1995ev}, particularly for the region with small scalar bare masses. It necessitates understanding how renormalized scalar masses manifest in the region equipped with features bearing similarity to the conventional symmetry breaking. Hence, in accordance with current understanding of Higgs mechanism, ground state of the theory becomes a natural choice for selecting the region in the parameter space.
\par
In this paper, the previously explored Wick Cutkosky's model \cite{Mufti:2018xqq} is further studied with the interaction coupling of different orders upto the value of 1.5 in GeVs. The approach of Dyson Schwinger equations (DSEs) \cite{Swanson:2010pw,Roberts:1994dr} is used for the current investigation. In the Lagrangian, no self interactions are considered for either Higgs or the scalar field \footnote{Throughout this paper, Higgs is referred to the doublet complex scalar field and scalar field is reserved for scalar singlet field, and Higgs bar is referred to $h^{\dagger}$.}. Since four point interactions for both fields can be generated by the 3-point Yukawa interaction term, for example using Feynman's box diagrams, Yukawa 3-point interaction is preferred over self interactions.
\par
An implicit assumption throughout the studies is that Wick Cutkosky model is a non-trivial theory. Though, the $\phi^{4}$ theory is found to be a trivial theory \cite{Jora:2015yga,Beg:1984yh,Aizenman:1981zz,Dashen:1983ts,Wolff:2009ke,Weisz:2010xx,Siefert:2014ela,Hogervorst:2011zw}, it is not conclusively established yet that adding further particles and their interactions renders the theory trivial. As for the case of Higgs, which is also a (complex doublet) scalar field, its interactions with gauge bosons \cite{Maas:2013aia} have not been found to render the theory trivial, which supports the assumption mentioned above.
\par
There are two related papers, one considering a larger parameter space \cite{tajdar:2018lat1} using a different approach \cite{Ruthe:2008rut}, and the second one \cite{tajdar:2018yukl2d} considering phenomenological aspects of the theory. There is also a paper \cite{tajdar:2018wcdyn} which addresses dynamic mass generation in the theory which is a further continuation of studies of the same model.
\section{Technical Details}
A significant part of technical details are also reported somewhere else \cite{Mufti:2018xqq} and are included here for the sake of self sufficiency. Euclidean version of the Lagrangian is given by 
\begin{equation} \label{Lagrangian}
L= \delta^{\mu \nu} \partial_{\mu} h^{\dagger} \partial_{\nu} h + m_{h}^{2} h^{\dagger} h + \frac{1}{2} \delta^{\mu \nu} \partial_{\mu} \phi \partial_{\nu} \phi + \frac{1}{2} m_{s}^{2} \phi^{2}+ \lambda  \phi h^{\dagger} h
\end{equation}
with Higgs fields (h) with SU(2) symmetry and $\phi$ a real scalar singlet field. $\lambda$ is the three point interaction coupling and has positive definite values. All higher order interactions are kept from the Lagrangian, including the four point self interaction for both fields in favor of a three point interaction vertex term \cite{Mufti:2018xqq}, as also mentioned above. The approach of DSEs \cite{Swanson:2010pw,Roberts:1994dr} is used to extract correlation functions, from which renormalized masses are extracted. Dyson Schwinger equations for propagators, $H^{ij}(p)$ for Higgs and $S(p)$ for scalar singlet fields, in momentum space are given by
\begin{equation} \label{hpr1:dse}
 H^{ij}(p)^{-1} = \delta^{ij} (\ p^{2} + m_{h}^{2} )\  + 2 \lambda \int \frac{d^{4}q}{(2\pi)^{4}} S(q) \Gamma^{ik}(-p,p-q,q) H^{kj}(q-p)
\end{equation}
\begin{equation} \label{spr1:dse}
 S(p) ^{-1}  = p^{2} + m_{s}^{2} + 2 \lambda  \int \frac{d^{4}q}{(2\pi)^{4}} H^{ik}(q) \Gamma^{kl}(q,p-q,-p) H^{li}(q-p)
\end{equation}
with $\Gamma^{kl}(u,v,w)$ being the three point Yukawa interaction vertex of Higgs, Higgs bar, and scalar fields with momentum u, v, and w, respectively. Higgs and Higgs bar fields have indices k and $\l$, respectively. The renormalization conditions for the propagators \cite{Roberts:1994dr} are
\begin{equation} \label{hpr:ren_condition}
H^{ij}(p)  |_{p^{2}=m_{h}^{2}}= \frac{\delta ^{ij}}{p^{2}+m_{h}^{2}} |_{p^{2}=m_{h}^{2}} 
\end{equation}
\begin{equation} \label{spr:ren_condition}
S(p)  |_{p^{2}=m_{s}^{2}}= \frac{1}{p^{2}+m_{s}^{2}} |_{p^{2}=m_{s}^{2}}
\end{equation}
The momentum setting is such that the Higgs and scalar propagators have momenta normal to each other, while the other Higgs propagator keeps the momentum value required for the four momentum conservation.
\par
The starting expressions of correlation functions, propagators and 3 point vertex in our case, are set to their tree level expressions. For every update of vertex and Higgs propagator, Newton Raphson's method is implemented locally while scalar propagators are calculated directly from Higgs propagators and the vertex between Higgs and scalar field using equation \ref{spr1:dse}. Hence, the propagators are updated or calculated directly from a DSE under the boundary conditions for them. The vertex is update to numerically evolve to the point that the DSEs \ref{hpr1:dse} and \ref{spr1:dse} are satisfied within the preselected size of local error. Uniqueness of solutions is implicitly assumed.
\par
As there are three unknown correlation functions, a common approach is to use three DSEs, two for the propagators here and a DSE for the vertex \cite{Swanson:2010pw,Roberts:1994dr,Rivers:1987hi}. Introduction of a DSE for vertex introduces further higher correlation functions depending on the theory, which eventually sets off a never ending tower of equations. This problem is cured by introducing truncations and modeling higher correlation functions in a number of ways \cite{Roberts:1994dr}. However, the resulting correlation functions may have model or truncation dependence. It is not unusual to compare results from DSE computations with those from other approaches, such as the method of lattice simulations \cite{Ruthe:2008rut} but this approach also has limitations due to the fact that almost all non-perturbative approaches \cite{Rivers:1987hi,Ruthe:2008rut} have their own limitations. On the contrary, the current investigation employs only the DSEs for field propagators while the renormalization conditions are meant to serve as restraints on all the three correlation functions in such a way that they satisfy the two DSEs. Hence, no assumptions or truncations was required for such an approach. The only additional constraint, which is only meant to control the numerical fluctuations, is implemented by requiring that local fluctuations never exceed beyond an order of magnitude in (Euclidean) space-time.
\par
In equation \ref{Lagrangian}, following standard literature \cite{Peskin:1995ev,Kaku:1993ym}, the potential term in the model is given by
\begin{equation} \label{potential-term}
U(\phi,h^{i}) = m_{h}^{2} h^{\dagger} h + \frac{1}{2} m_{s}^{2} \phi^{2}+ \lambda  \phi h^{\dagger} h
\end{equation}
As the model contains two fields, the lowest of potential should be dictated by both of them. It leads to the following equation. 
\begin{equation} \label{potential2a}
m_{s}^{2}=\frac{\nu^{2}}{m_{h}^{2}}   \lambda^{2}
\end{equation}
which also implies that
\begin{equation} \label{potential2b}
m_{s}^{2} \propto   \lambda^{2}
\end{equation}
where $\nu$ is the vacuum expectation value (vev) of Higgs field.
\par
There are two interesting features of equation \ref{potential2a}. First, there is no role of the vacuum expectation value from $\phi$ field, and only the knowledge of Yukawa coupling, Higgs' vev and Higgs' tree level mass is sufficient to determine the tree level mass of $\phi$ field. However, it is important to note that $\lambda$ is a three point interaction coupling instead of the four point self interaction coupling. As these two types of interactions have different couplings, it is not straight forward to extract value of $\lambda$ from the four point coupling value. It was circumvented in previous work \cite{Mufti:2018xqq} by introducing a non-dynamic parameter in the theory and fix it from the start. However, this approach may not be helpful here as the objective is to search for regions in parameter space with specific features of the model itself.
\par
Secondly, if there is no significant beyond tree level contributions and tree level mass of Higgs remains essentially the dominant contributor in Higgs renormalized mass, the proportionality may remain established for renormalized masses in the theory. This expectation is reasonable as it has already been observed in this model \cite{Mufti:2018xqq} as well as in other studies \cite{Gies:2017zwf}.
\par
In this investigation, $10^{-5} \leq \lambda \leq 1.5$ in GeVs are selected to examine the relationship \ref{potential2a}. As the theory has also been found to be cutoff dependent \cite{Mufti:2018xqq}, every considered point in the parameter space is studied using cutoff values at 10 TeV, 20 TeV, and 30 TeV. Maximum acceptable local error for correlation functions is set to $10^{-18}$. As mentioned above, Higgs bare mass is set at 125.09 GeV throughout the studies.
\par
Gaussian quadrature algorithm is used for numerical integrations.
\section{Propagators}
\subsection{Scalar Propagators}
\begin{figure}
\centering
\parbox{0.92\linewidth}
{
\vspace{-2cm}
\includegraphics[width=1.0\linewidth, angle=0.0]{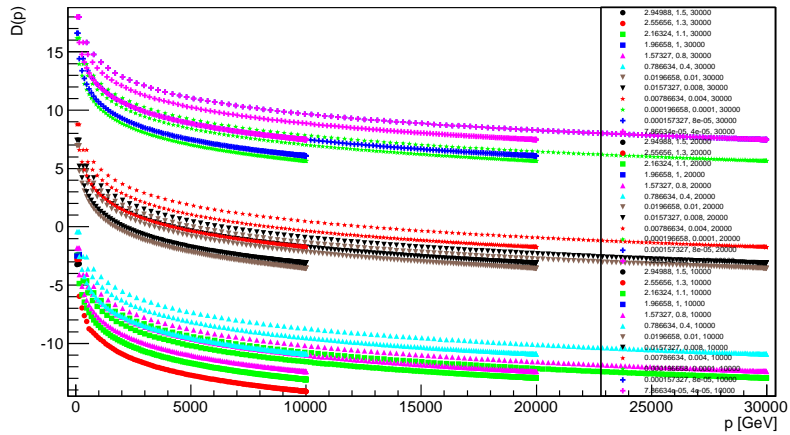}
\caption{\label{yukl-sprs1} Scalar propagators (on logarithmic scale) for different bare coupling and cutoff values are plotted.}
}
\end{figure}
\begin{figure}
\vspace{-2cm}
\centering
\parbox{0.9\linewidth}
{
\includegraphics[width=\linewidth]{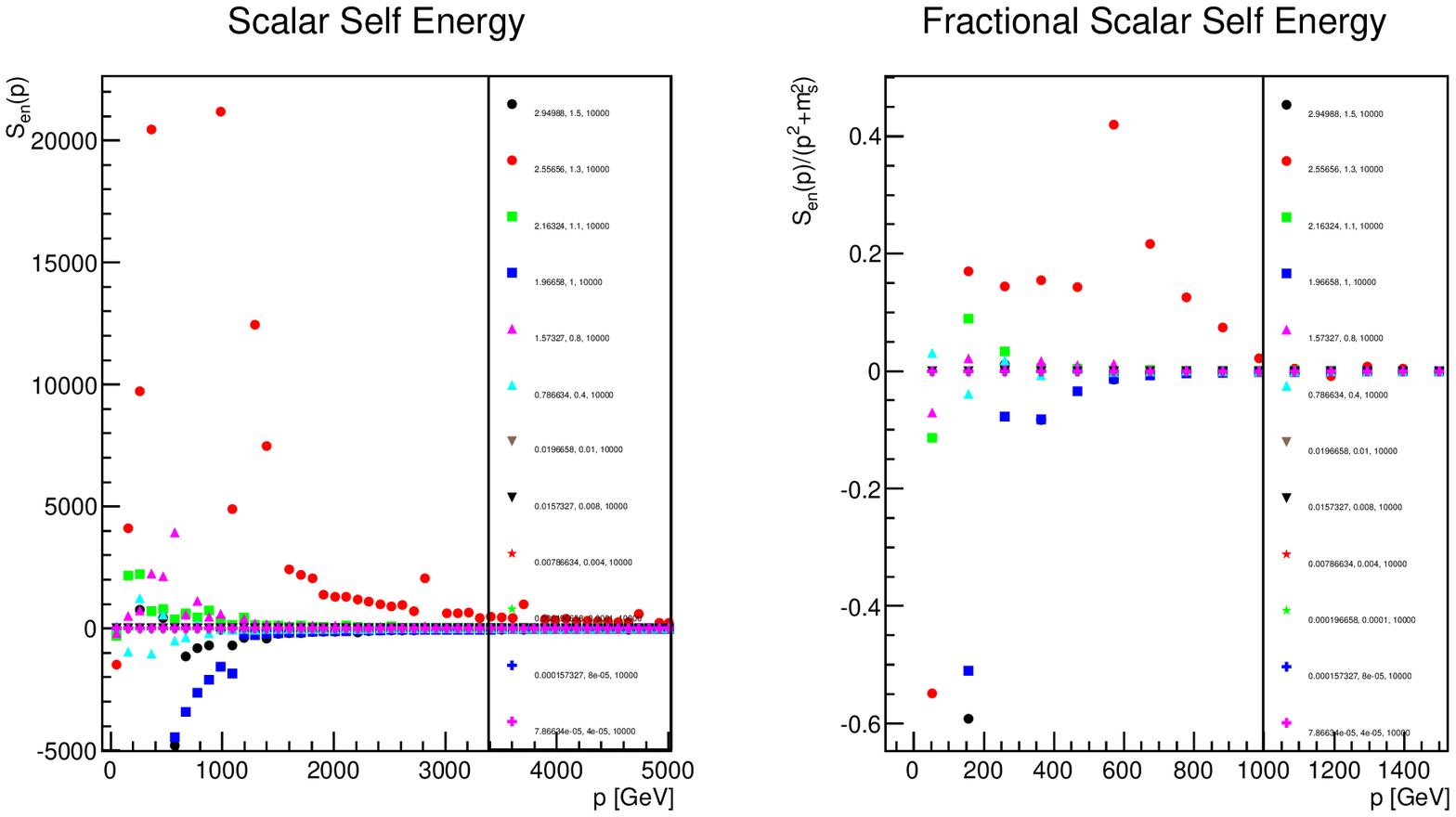}
\caption{\label{yukl-ssen1-10} Scalar self energy for 10 TeV cutoff are plotted for different bare coupling values.}
}
\centering
\parbox{0.9\linewidth}
{
\centering
\includegraphics[width=\linewidth]{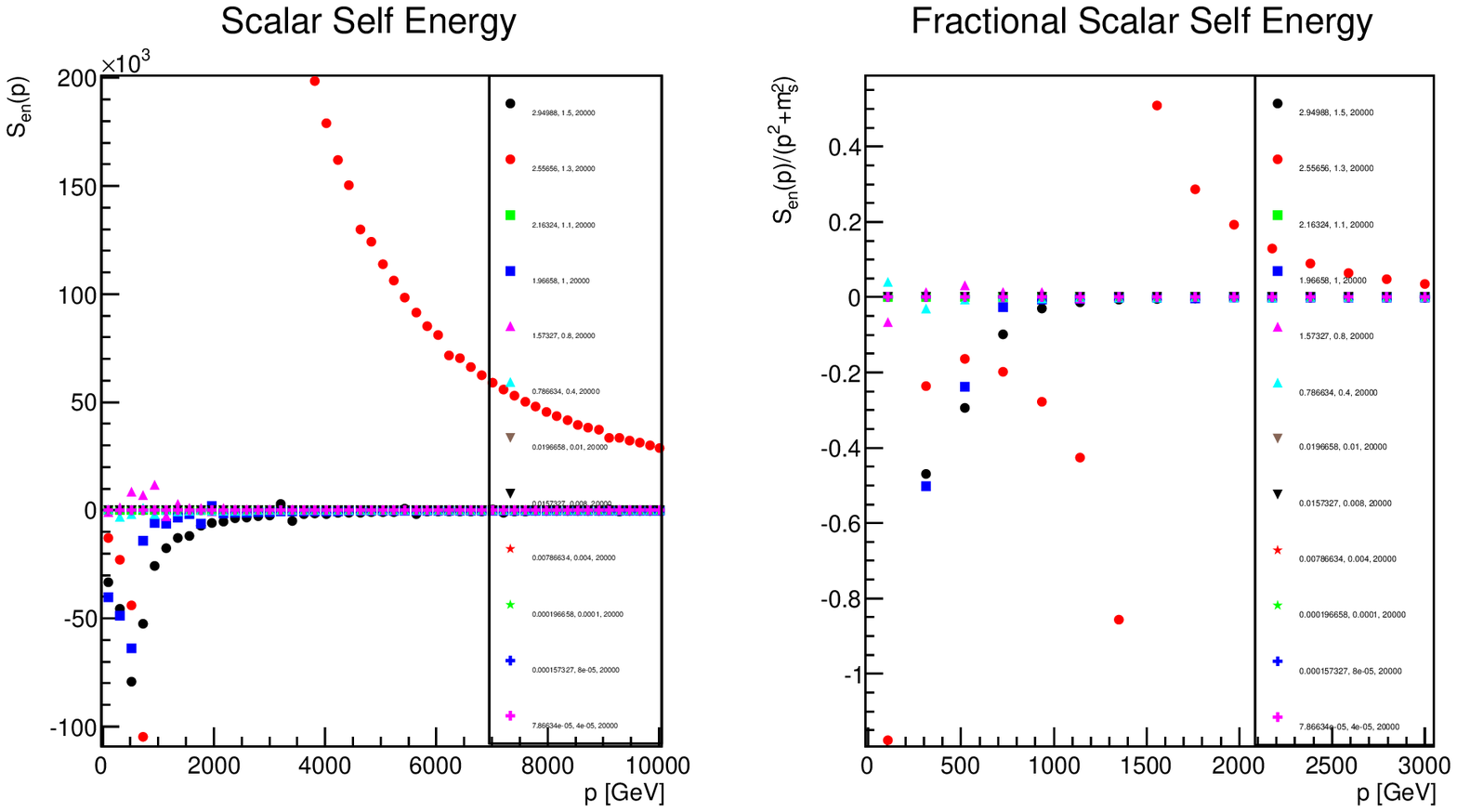}
\caption{\label{yukl-ssen1-20} Scalar self energy for 20 TeV cutoff are plotted for different bare coupling values.}
}
\centering
\parbox{0.9\linewidth}
{
\centering
\includegraphics[width=\linewidth]{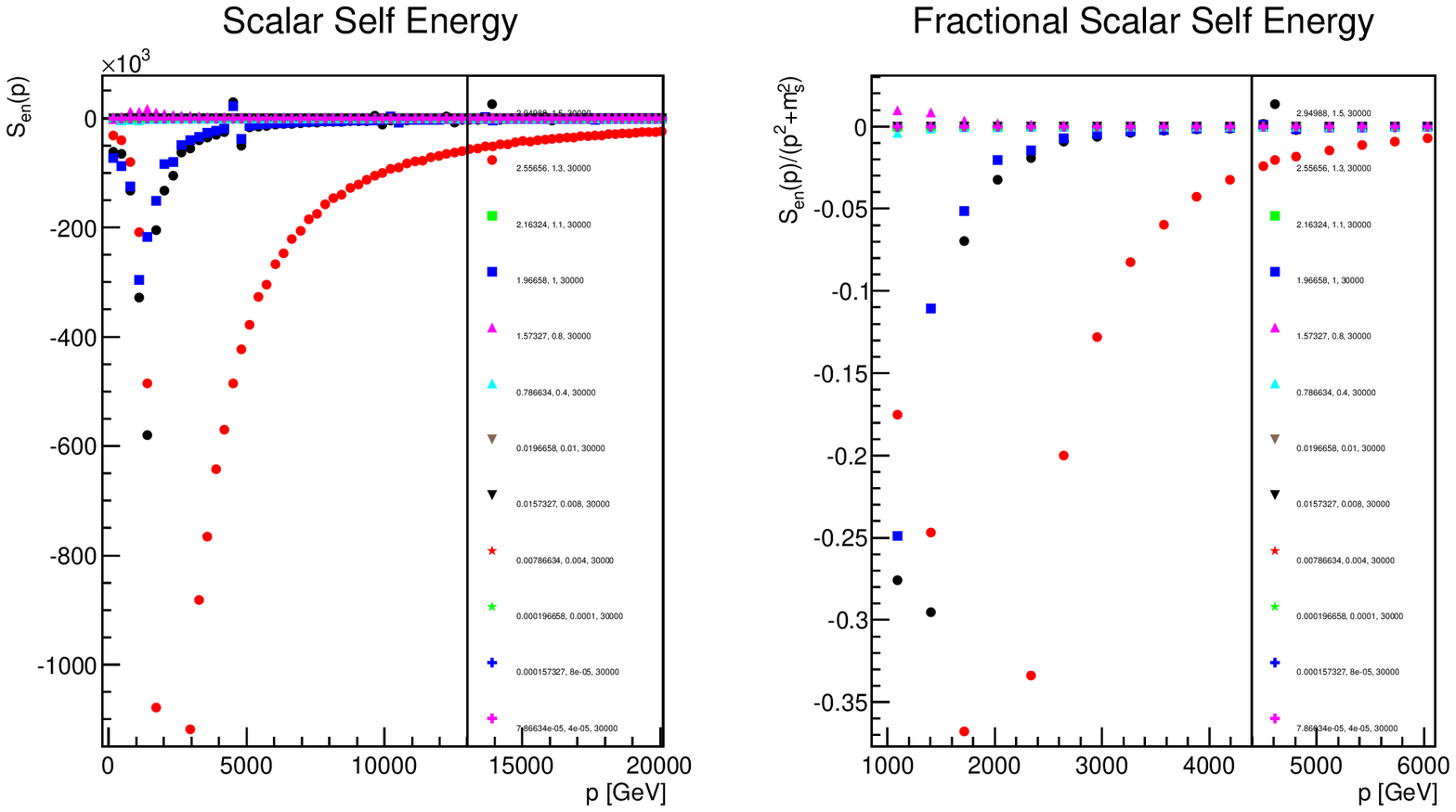}
\caption{\label{yukl-ssen1-30} Scalar self energy for 30 TeV cutoff are plotted for different bare coupling values.}
}
\end{figure}

Propagators are the simplest non-zero correlation functions in the theory. They are calculated under the renormalization conditions mentioned in \ref{hpr:ren_condition}-\ref{spr:ren_condition} for Higgs and scalar singlet fields, respectively.
\par
Scalar propagators are shown in figure \ref{yukl-sprs1} for different bare couplings and cutoff values. In comparison to the previous investigation of the theory \cite{Mufti:2018xqq} carried out in a different region of the parameter space, scalar propagators manifest themselves markedly differently. On one hand, in ultraviolet region they have the same characteristic of multiplicatively renormalized propagators and considerable difference than the tree level propagators towards the infrared region. However, they do not accumulate in groups in this region of parameter space which was observed before \cite{Mufti:2018xqq}. Since, this behavior persists even in the ultraviolet region, it suggests that the theory may have relatively wider spectrum of renormalized scalar masses which was not as vividly observed in the previous investigation \cite{Mufti:2018xqq}. Furthermore, a general trend is greater the coupling (hence greater the bare masses due to equations \ref{potential2a} and \ref{potential2b}), more suppressed the ultraviolet end appears. A peculiar observation is that for the two higher cutoff values (20 and 30 TeVs), scalar propagators are less different than for 10 TeV cutoff. It suggests that, should this trend persist, at a cutoff not far from 30 TeV (say 100 TeV) the propagators may become extremely similar to each other, hence may have very similar masses. If this is true, the choice for scalar propagators in selecting masses for the theory may also reduce drastically.
\par
The scalar self energy terms, see figures \ref{yukl-ssen1-10}-\ref{yukl-ssen1-30}, are one of the most interesting results here, particularly in comparison to \cite{Mufti:2018xqq}. The immediate observation is immense negative contributions from them, beside some positive contributions for a number of bare couplings. Most of the bare masses during this investigation are less than 5 GeV. For such small bare masses, even in tree level contribution most of the contributions come from momentum rather than the bare mass. Hence, in this region of parameter space it is much easier for the renormalized masses to receive significant contributions in comparison to the bare mass values. As these contributions appear infrared enhanced, negative contributions may easily change the overall sign of squared scalar masses. Hence, the speculations made in \cite{Mufti:2018xqq} regarding a pole appearing in scalar propagators due to negative renormalized squared scalar masses is strongly favored in the considered region of parameter space. Scalar self energy terms are also in favor of sufficiency of a cutoff not far above 30 TeV as they behave relatively more similarly for 20 TeV and 30 TeV in comparison to 10 TeV cutoff.
\subsection{Higgs Propagators}
\begin{figure}
\centering
\parbox{0.92\linewidth}
{
\vspace{-2cm}
\includegraphics[width=1.0\linewidth, angle=0.0]{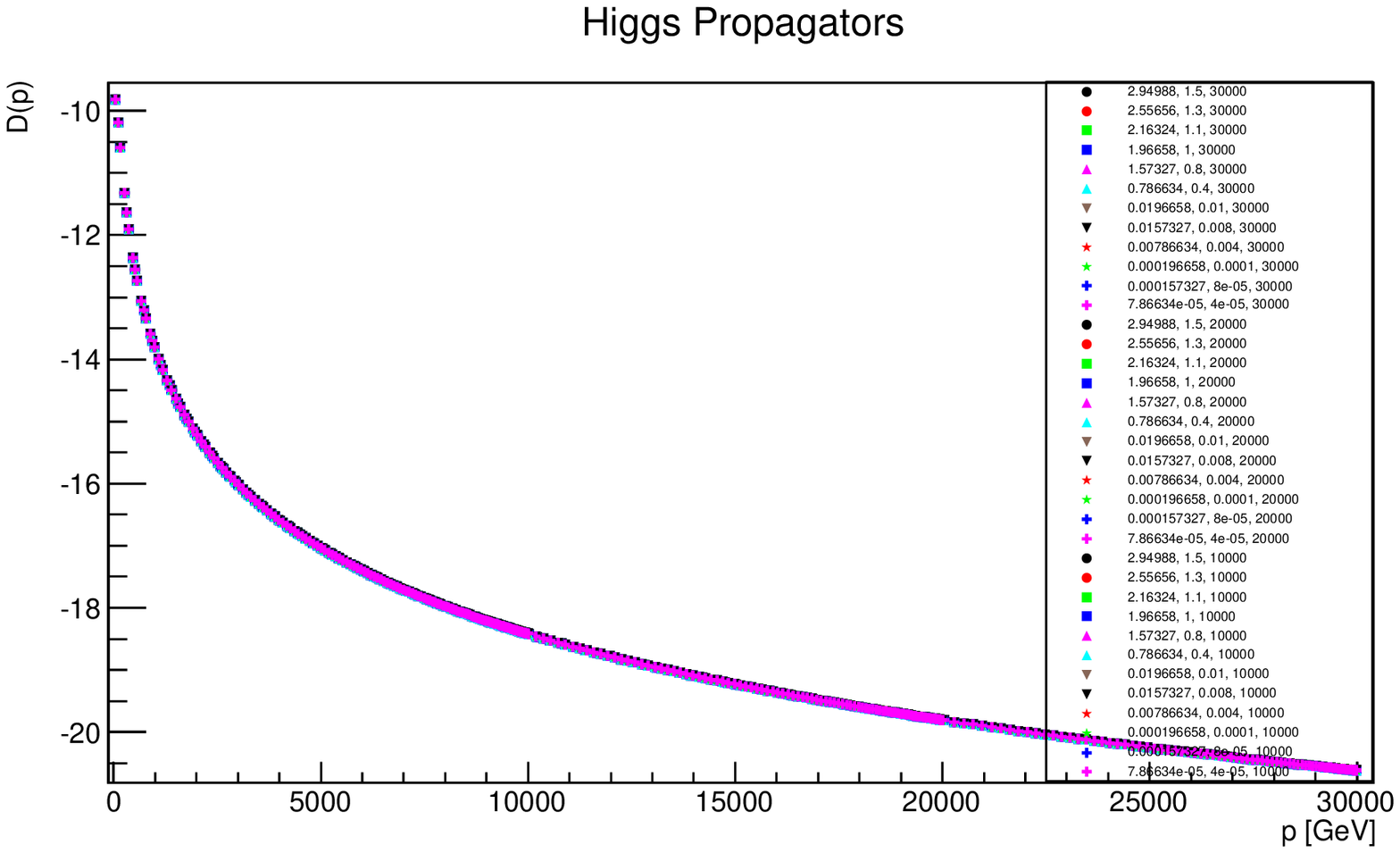}
\caption{\label{yukl-hprs1} Higgs propagators (on logarithmic scale) for different bare coupling and cutoff values are given.}
}
\end{figure}
\begin{figure}
\vspace{-2cm}
\centering
\parbox{0.9\linewidth}
{
\includegraphics[width=\linewidth]{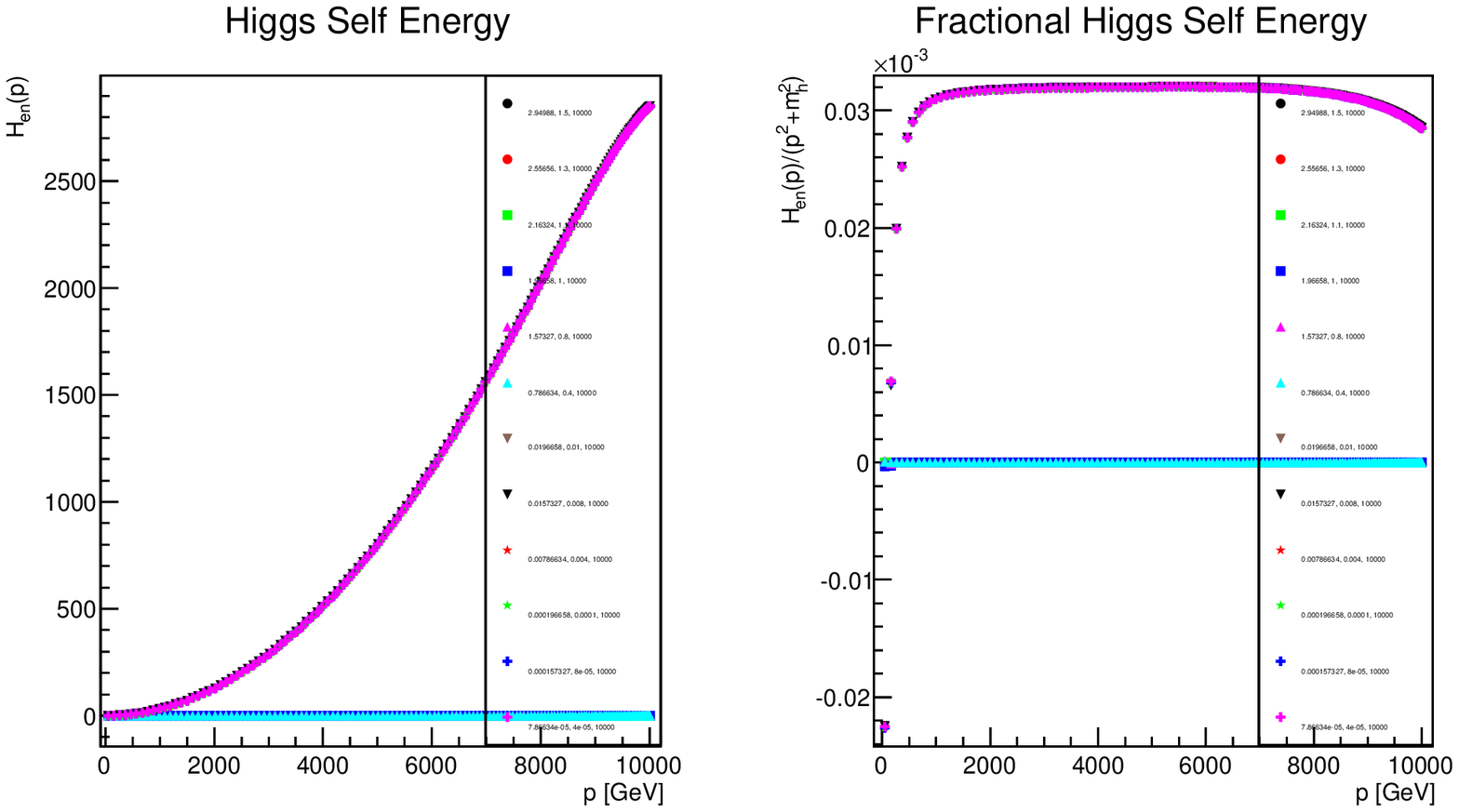}
\caption{\label{yukl-hsen1-10} Higgs self energy for 10 TeV cutoff are plotted for different bare coupling values.}
}
\centering
\parbox{0.9\linewidth}
{
\centering
\includegraphics[width=\linewidth]{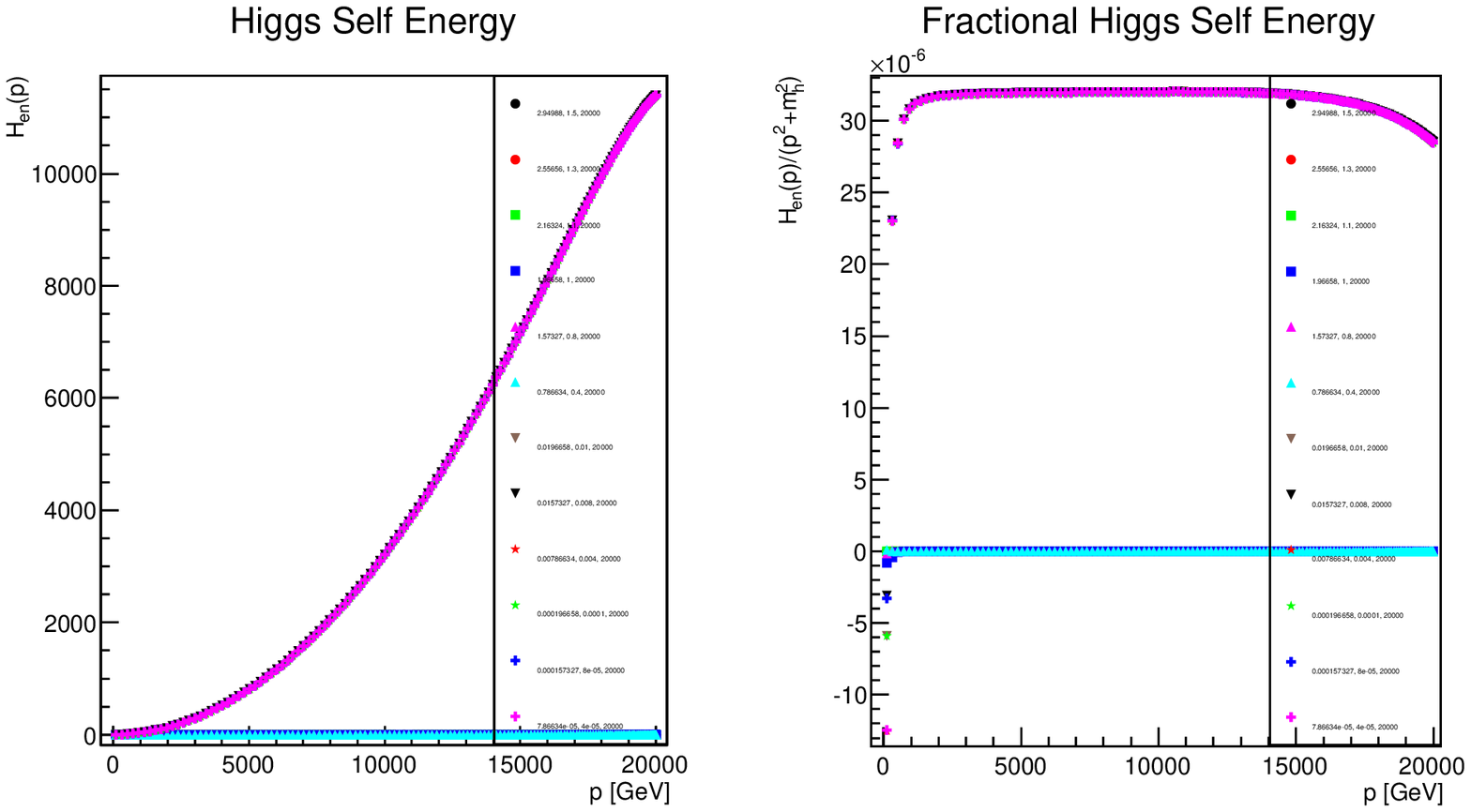}
\caption{\label{yukl-hsen1-20} Higgs self energy for 20 TeV cutoff are plotted for different bare coupling values.}
}
\centering
\parbox{0.9\linewidth}
{
\centering
\includegraphics[width=\linewidth]{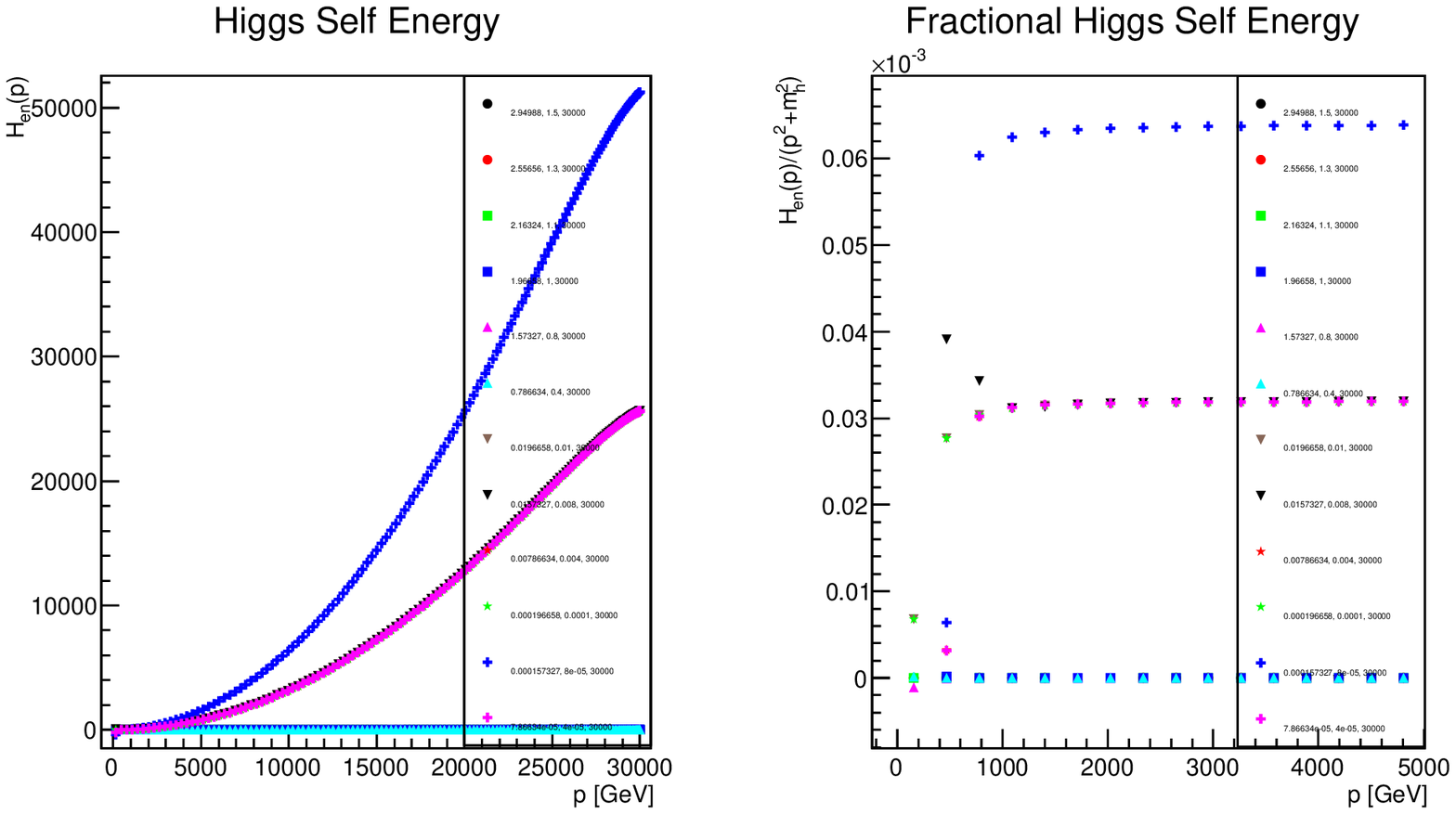}
\caption{\label{yukl-hsen1-30} Higgs self energy for 30 TeV cutoff are plotted for different bare coupling values.}
}
\end{figure}
Higgs propagators are shown in figure \ref{yukl-hprs1} \footnote{All Higgs propagators are found to be quantitatively close, hence are not clearly visible separately. The difference in their behavior can be more clearly seen in their self energy terms.}. In comparison to the previous investigation \cite{Mufti:2018xqq}, in this region of parameter space Higgs propagators do not show remarkable (quantitative) resemblance. An immediate observation is that these propagators behave differently at both infrared as well as ultraviolet end. Hence, the tree structure of propagator does not seem to be as dominant contributor for all the considered parameters. However, the effects are significant for all the momentum values and, despite the presence of contributions beyond the tree level structure, qualitatively Higgs propagators manifest similar behavior, see figure \ref{yukl-hprs1}. 
\par
The quantitative dissimilarities can be seen in the Higgs self energy terms, shown in figures \ref{yukl-hsen1-10}-\ref{yukl-hsen1-30} (left). There are two distinct behaviors of Higgs self energies. Either it manifests as an almost constant function over a wide range of momentum, or it is an increasing function over a wide range of momentum values. These increasing self energies contribute in suppressing their respective Higgs propagators for large momentum values. However, their contributions in comparison with the tree level structure for smaller cutoff values seem to start diminishing in the vicinity of ultraviolet end, see (right) figures \ref{yukl-hsen1-10}-\ref{yukl-hsen1-20} while for 30 TeV these relative contributions are qualitatively the same as the tree level, see (right) figure \ref{yukl-hsen1-30}. It implies that for higher cutoff values, Higgs propagators tend to take the form of a tree level structure upto a renormalization constant.
\par
A peculiar observation is the presence of negative contributions which are relatively more abundant than in the previously investigated region in parameter space \cite{Mufti:2018xqq}. However, these contributions are far less in magnitude to considerably effect the propagators either qualitatively or quantitatively.
\section{Renormalized Masses}
\begin{figure}
\centering
\parbox{1.0\linewidth}
{
\includegraphics[width=1.0\linewidth]{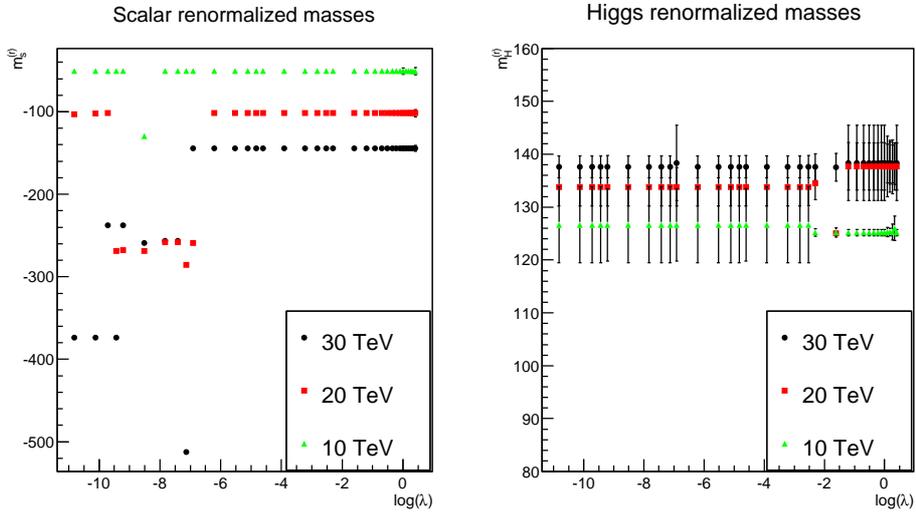}
\caption{\label{masses1} Renormalized masses in the model for different cutoff values are plotted. Sign of the masses are such that if the respective renormalized squared mass is positive (negative), the magnitude is shown with positive (negative) sign.}
}
\end{figure}
\begin{figure}
\centering
\parbox{1.0\linewidth}
{
\includegraphics[width=1.0\linewidth]{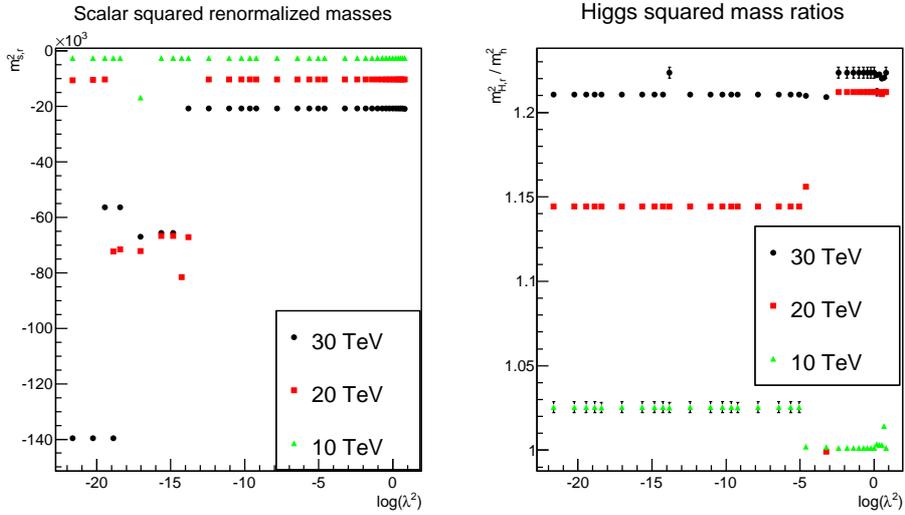}
\caption{\label{masses2} Scalar squared renormalized masses (left) and ratio of Higgs squared renormalized masses $m^{2}_{H},r$ and Higgs squared bare mass $m^{2}_{H}$ (right) for different cutoff values are plotted.}
}
\end{figure}
Once the propagators for Higgs and scalar fields are computed, it is relatively straight forward to extract renormalized masses using suitable polynomial fits. The following curve fitting is used for inverse propagators $D(p)^{-1}$ for both Higgs and scalar propagators.
\begin{equation} \label{fit-of-propagators}
D(p)^{-1} = \sum_{i=-l}^{i=l} a_{i} p^{2i}
\end{equation}
The results are shown in figures \ref{masses1} and \ref{masses2}.
\par
For Higgs renormalized masses, agreement of masses within $2\sigma$ for most of the values of bare coupling is observed, which is in fact an indicative of small corrections coming from self energy terms. It further substantiates understanding of Higgs masses explored previously in the same model \cite{Mufti:2018xqq} as well as some other similar models \cite{Gies:2017zwf}. As the previous study \cite{Mufti:2018xqq} belonged to scalar masses with entirely different order of magnitude, it supports the speculation that Higgs may have dominant contribution to its renormalized mass from the tree level term of Higgs propagator throughout the parameter space. The situation slightly changes for higher cutoff values as the self energy terms tend to contribute more. However, for higher cutoff values Higgs renormalized masses are found to be loosing dependence on coupling values, see figures \ref{masses1} (right) and \ref{masses2} (right), which suggests that beyond some higher cutoff value not far from 30 TeV (such as 100 TeV) the dependence on coupling values will practically be lost and the theory may generate a cutoff independent renormalized mass in the theory. As the figure \ref{masses2} (right) indicates, 3-point Yukawa interactions have a tendency to generate higher Higgs physical masses than what has been observed experimentally.
\par
The squared scalar renormalized masses (SSRM) have, however, very different manifestation compared to the previously explored region of parameter space \cite{Mufti:2018xqq}. An immediate observation is that SSRM are negative for most of the parameters while squared bare masses for both fields were kept positive throughout the investigation in an attempt to ease numerical implementation of renormalization scheme and to establish a common phase to start computations for entire studies, see figure \ref{masses2} (left). These negative values further support the speculation made previously \cite{Mufti:2018xqq} regarding the scalar field exhibiting Higgs mechanism-like behavior.
\par
A peculiar observation is that for low cutoff values, the SSRM do not receive significant coupling dependent contributions for most of the couplings, see figure \ref{masses2} (left). However, a non trivial dependence is observed over a certain region of coupling values as the cutoff is increased, see figures \ref{masses1} (left) and \ref{masses2} (left). The apparent trend is that as the cutoff value is increased, relatively wider region of coupling values contributes to the coupling dependent behavior which is qualitatively closer to the equation \ref{potential2a}, though one of difference is an additive constant.
\section{Three Point Vertex}
\begin{figure}
 \centering
 \parbox{0.47\linewidth}
 {
 \includegraphics[width=\linewidth]{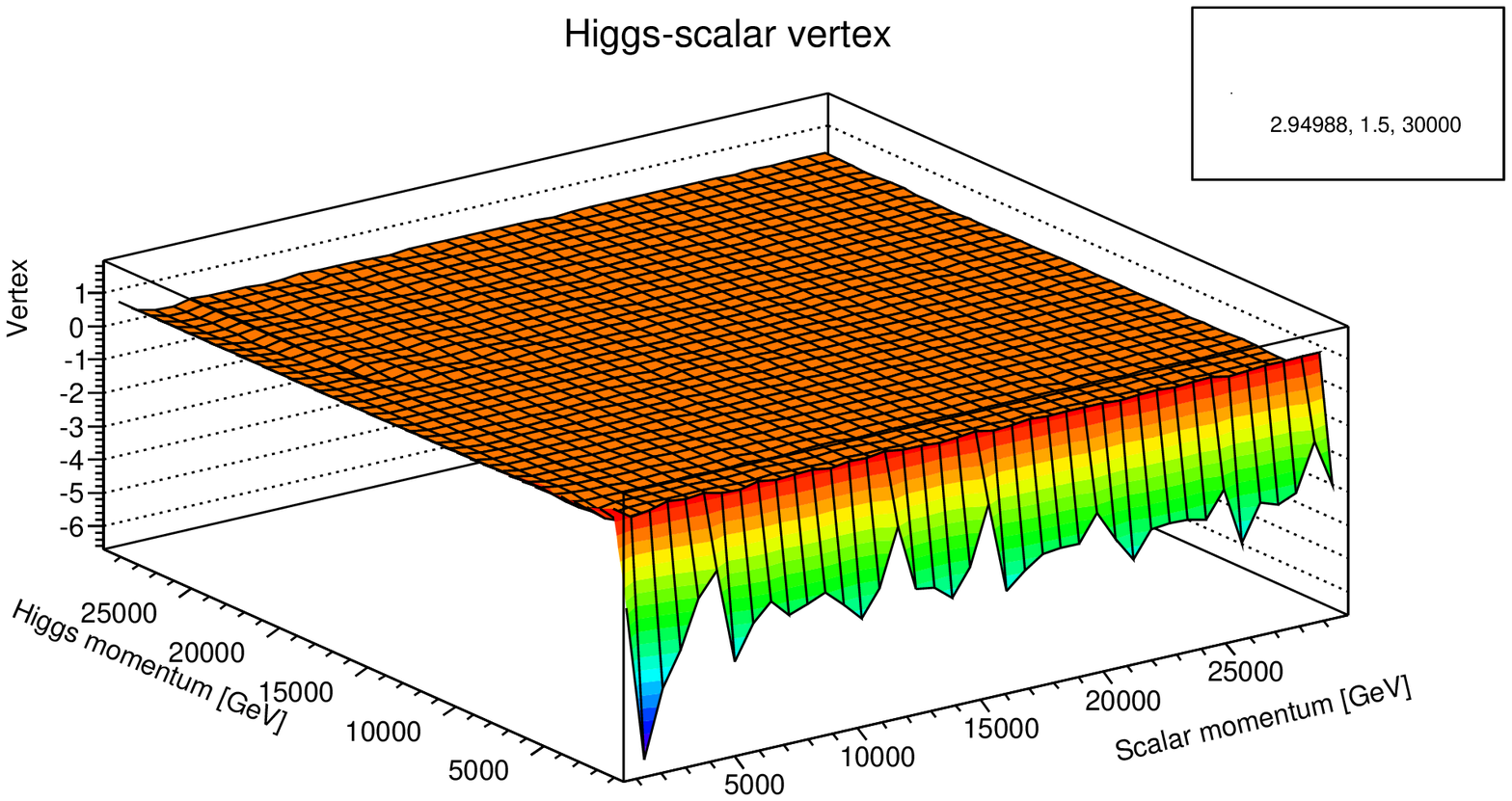}
 }
 \parbox{0.47\linewidth}
 {
 \includegraphics[width=\linewidth]{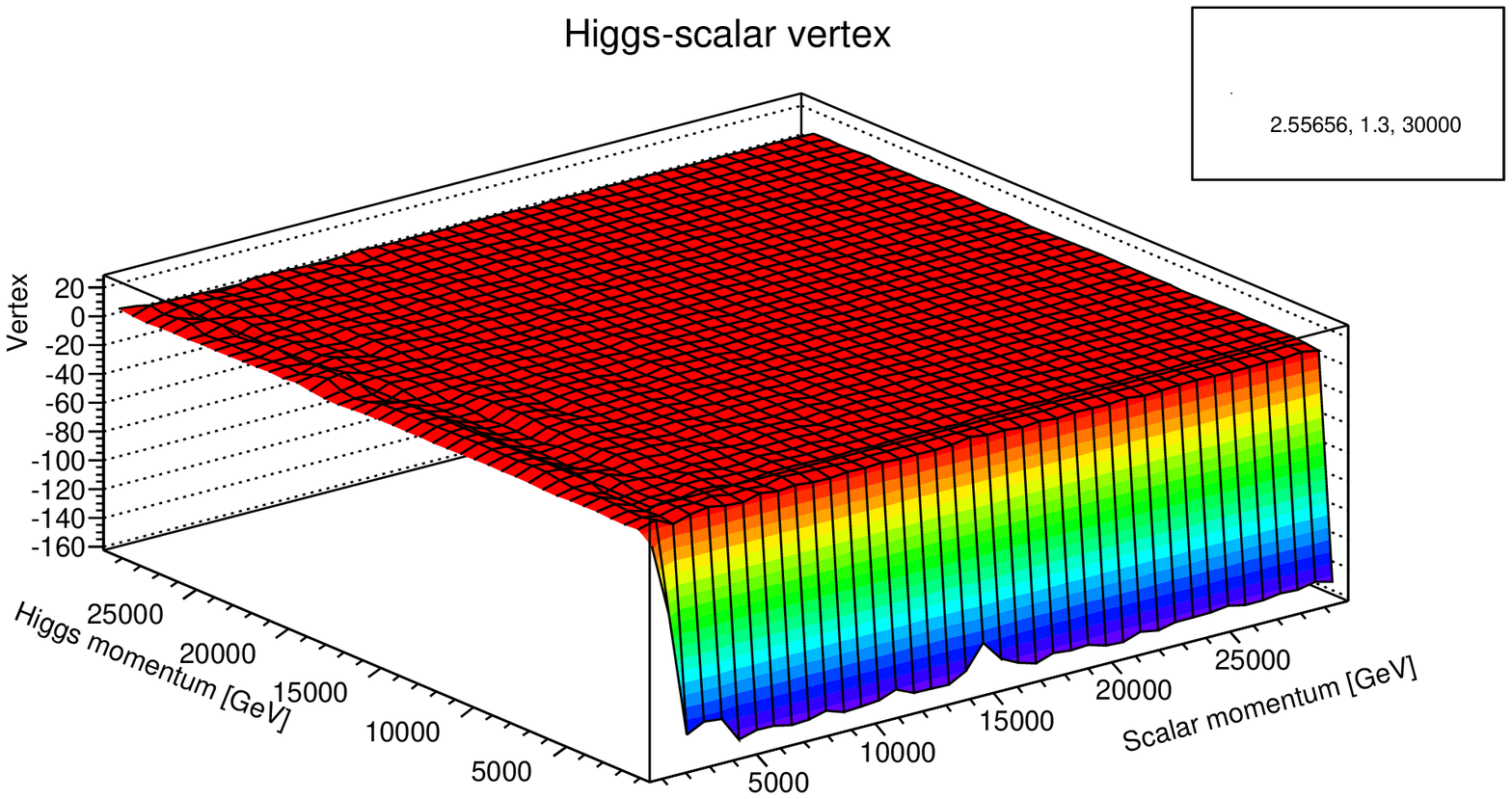}
 }
 \centering
 \parbox{0.47\linewidth}
 {
 \includegraphics[width=\linewidth]{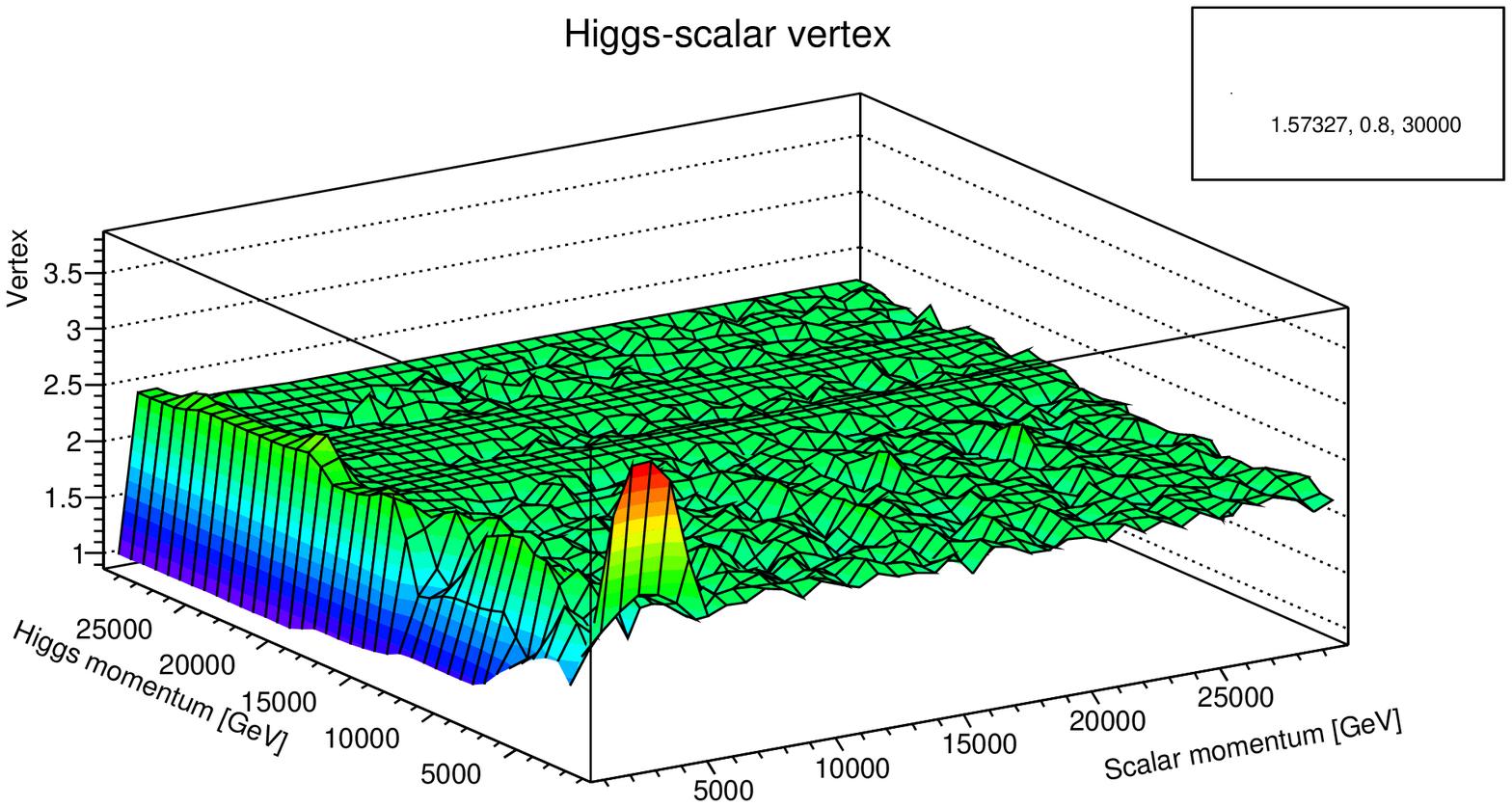}
 }
  \parbox{0.47\linewidth}
 {
 \includegraphics[width=\linewidth]{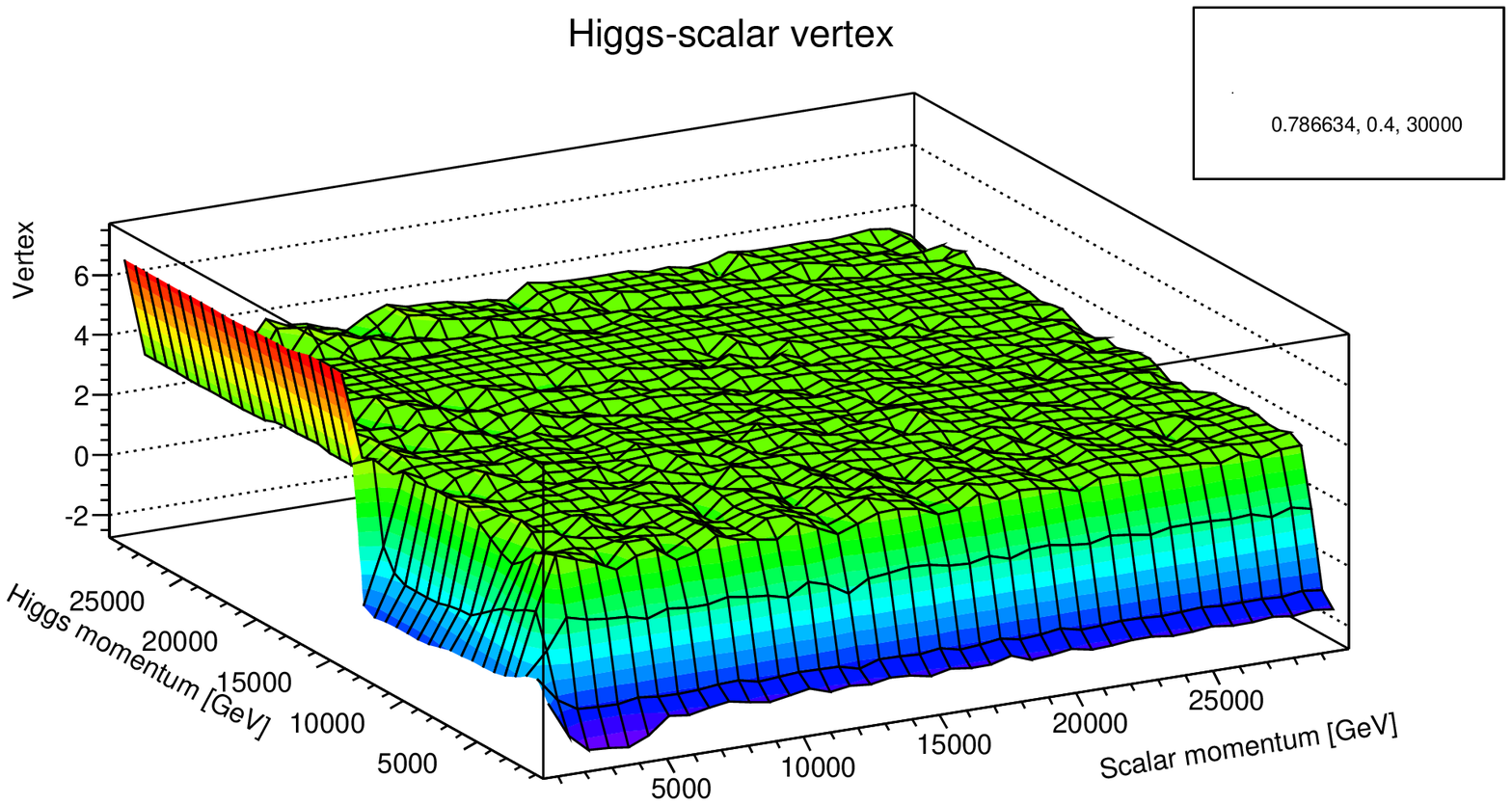}
 }
 \centering
 \parbox{0.47\linewidth}
 {
 \includegraphics[width=\linewidth]{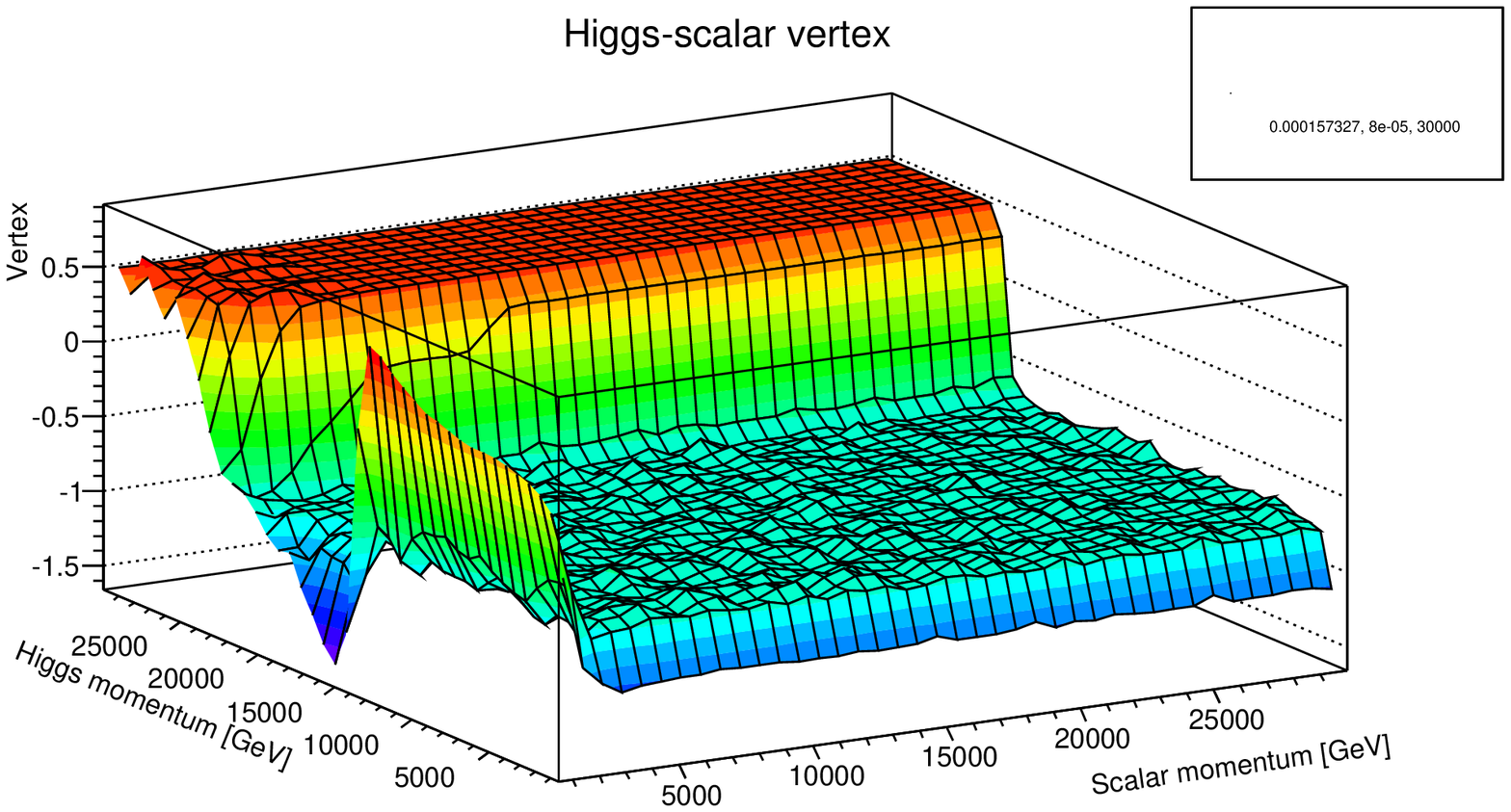}
 }
  \parbox{0.47\linewidth}
 {
 \includegraphics[width=\linewidth]{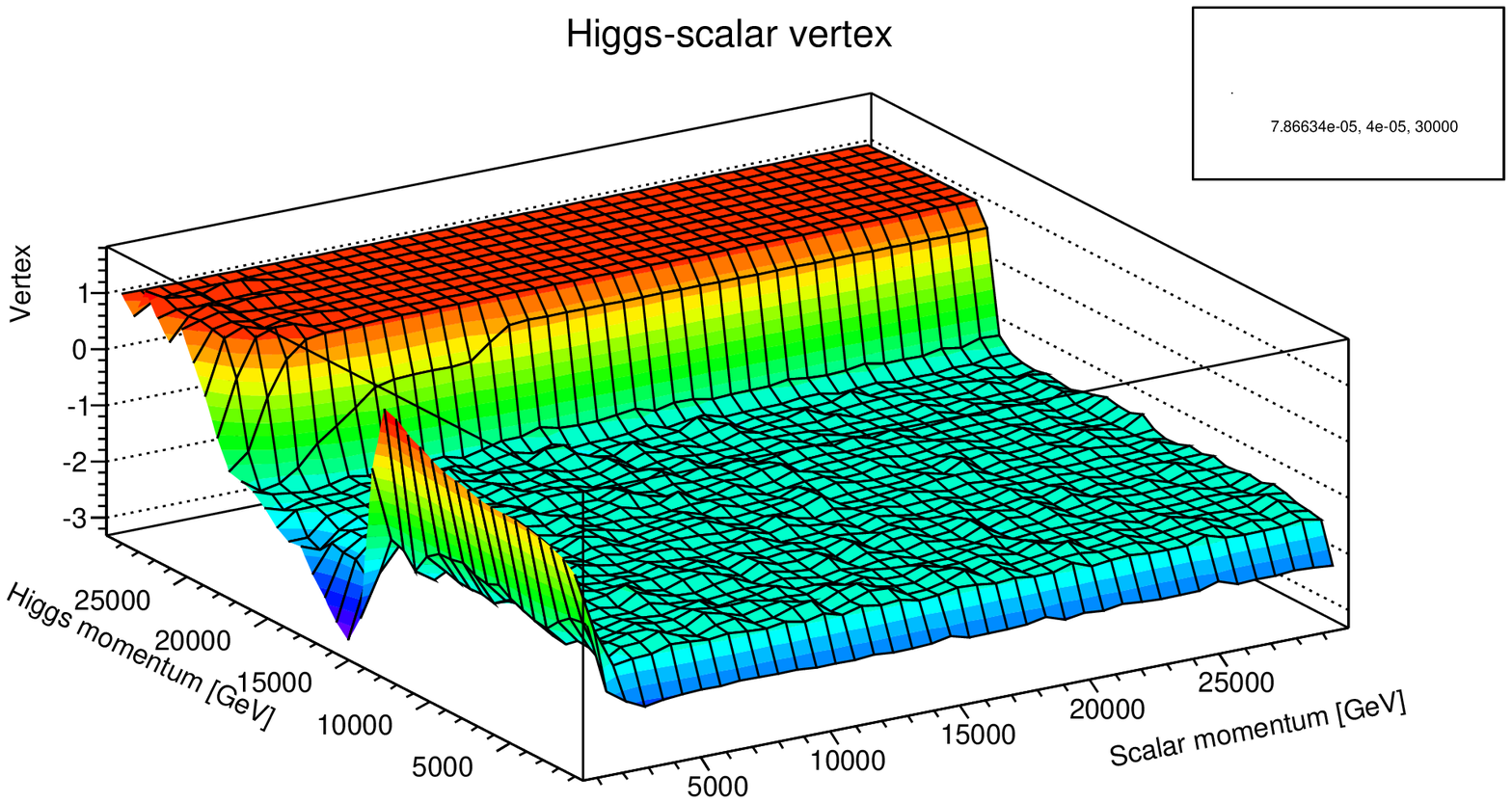}
 }
 \caption{\label{yukl-v-30} Higgs-scalar vertices for different bare couplings are plotted for 30 TeV cutoff value.}
\end{figure}
\begin{figure}
 \centering
 \parbox{0.47\linewidth}
 {
 \includegraphics[width=\linewidth]{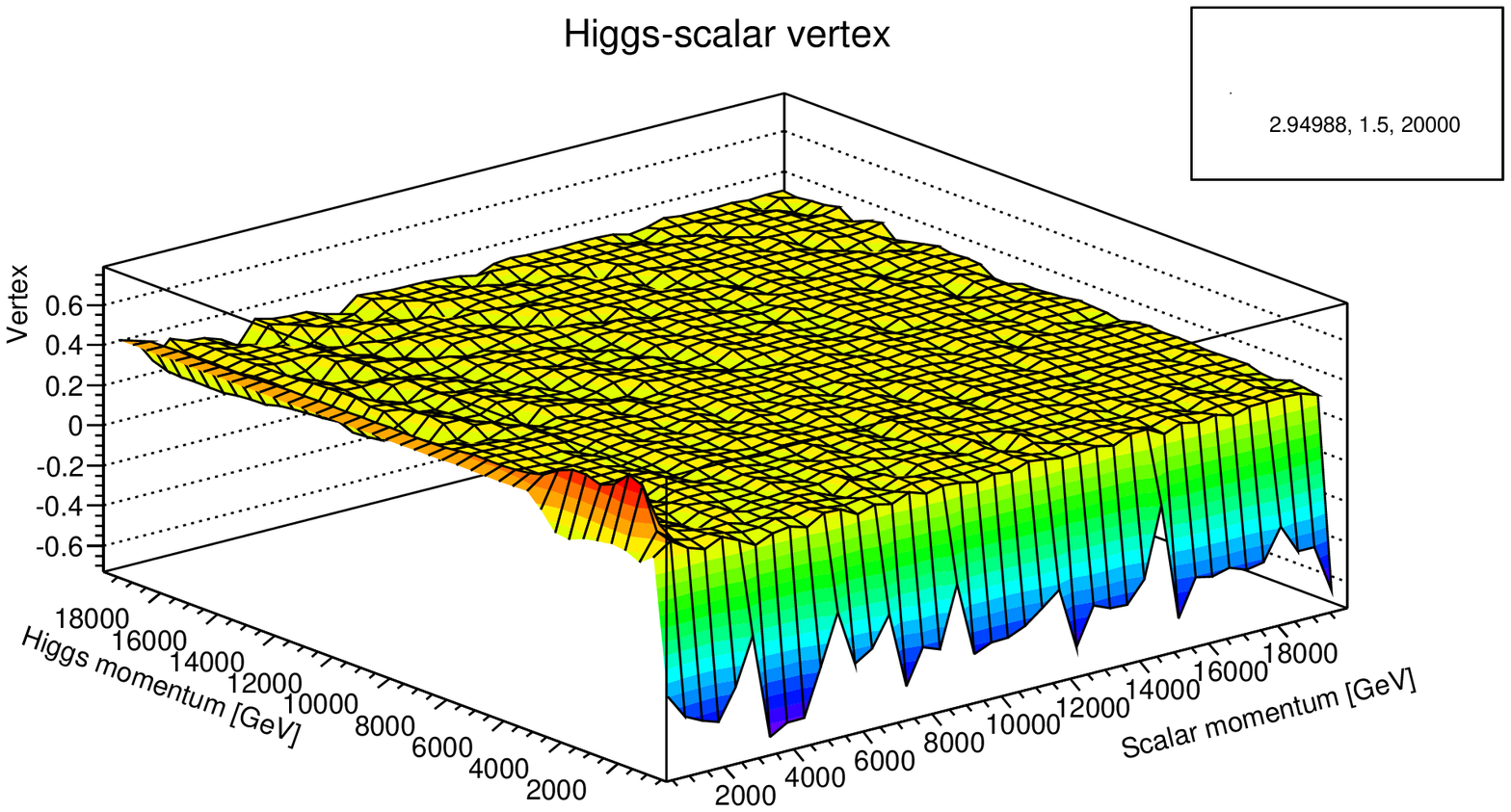}
 }
 \parbox{0.47\linewidth}
 {
 \includegraphics[width=\linewidth]{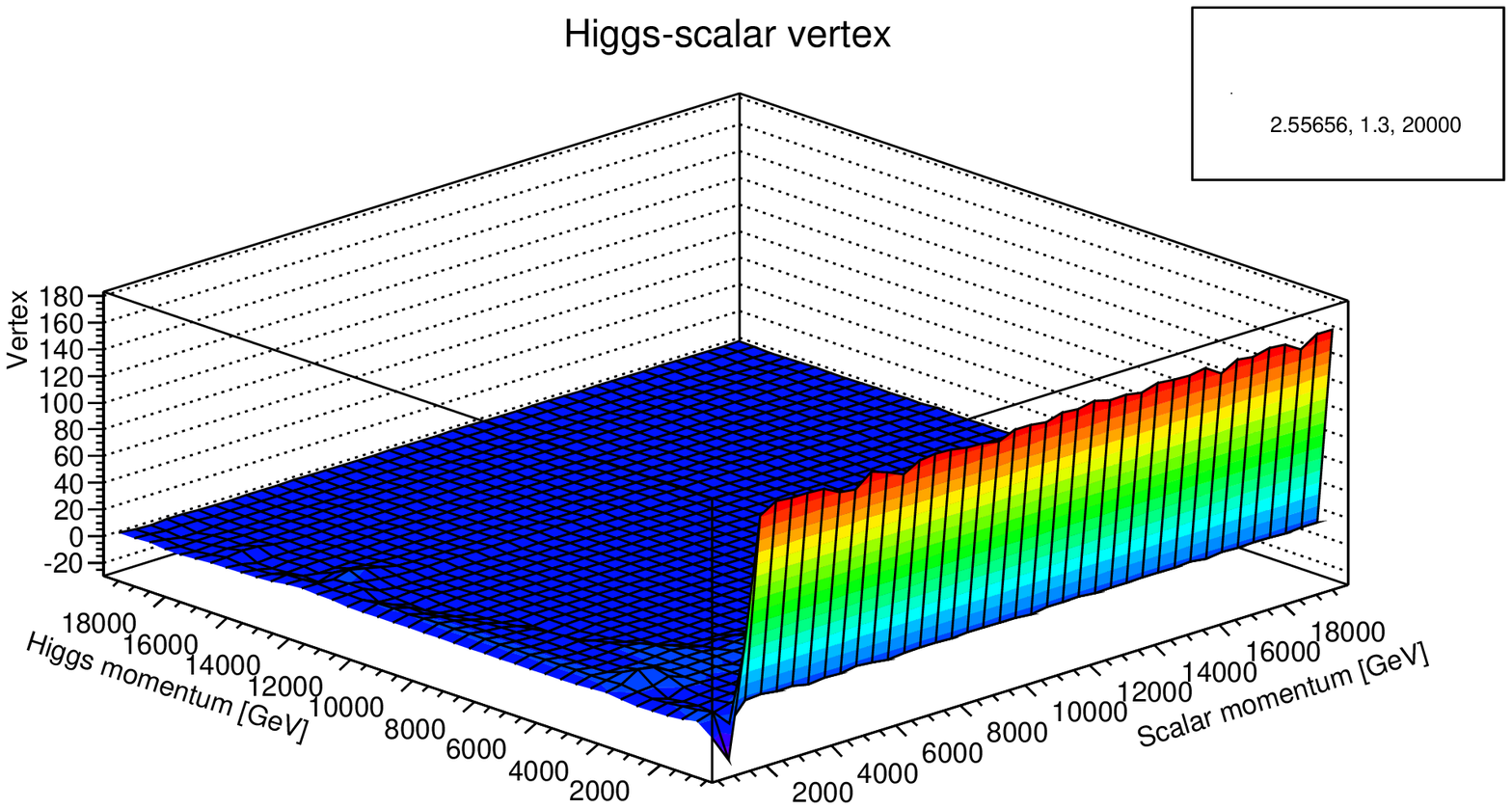}
 }
 \centering
 \parbox{0.47\linewidth}
 {
 \includegraphics[width=\linewidth]{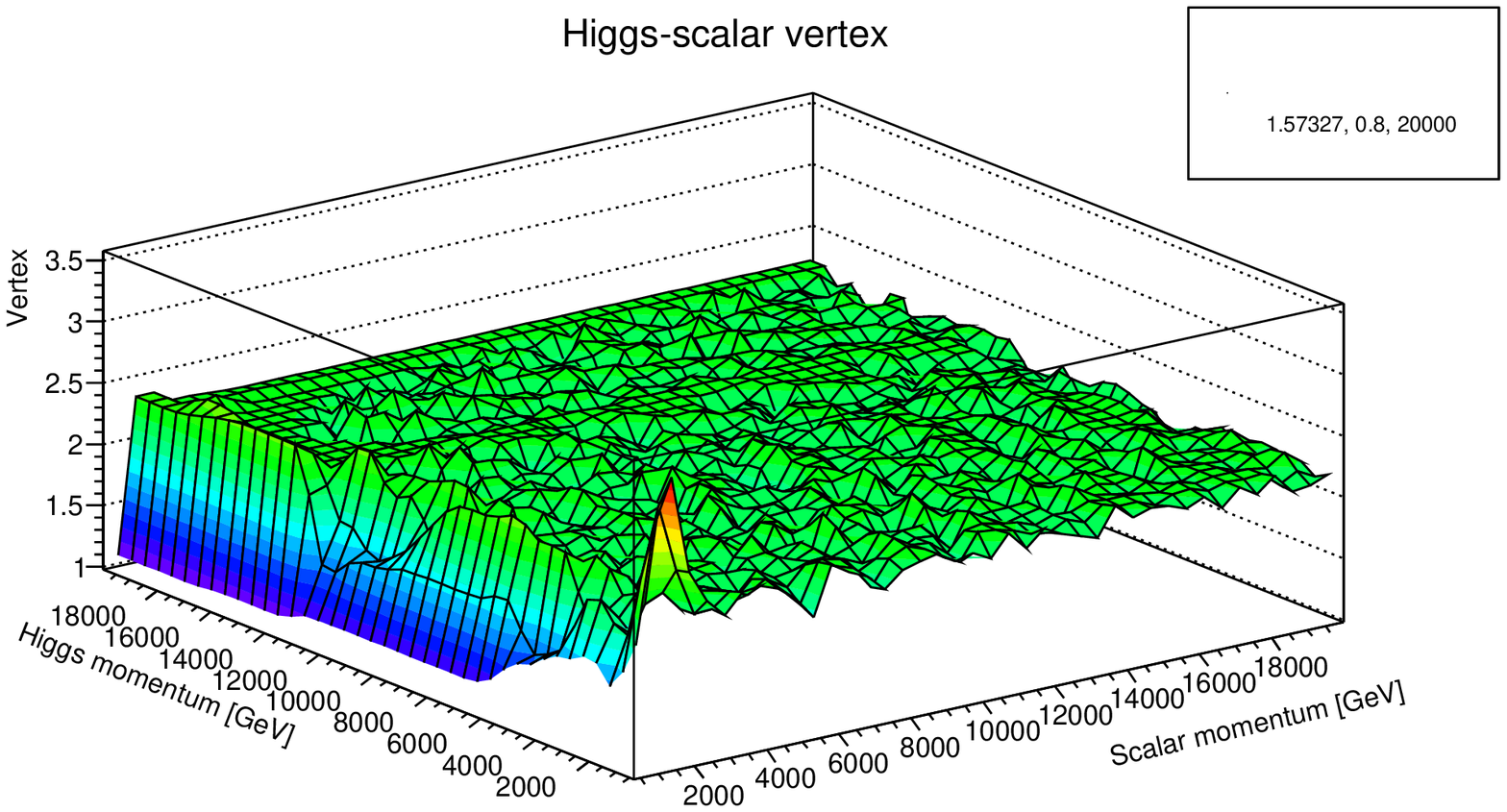}
 }
  \parbox{0.47\linewidth}
 {
 \includegraphics[width=\linewidth]{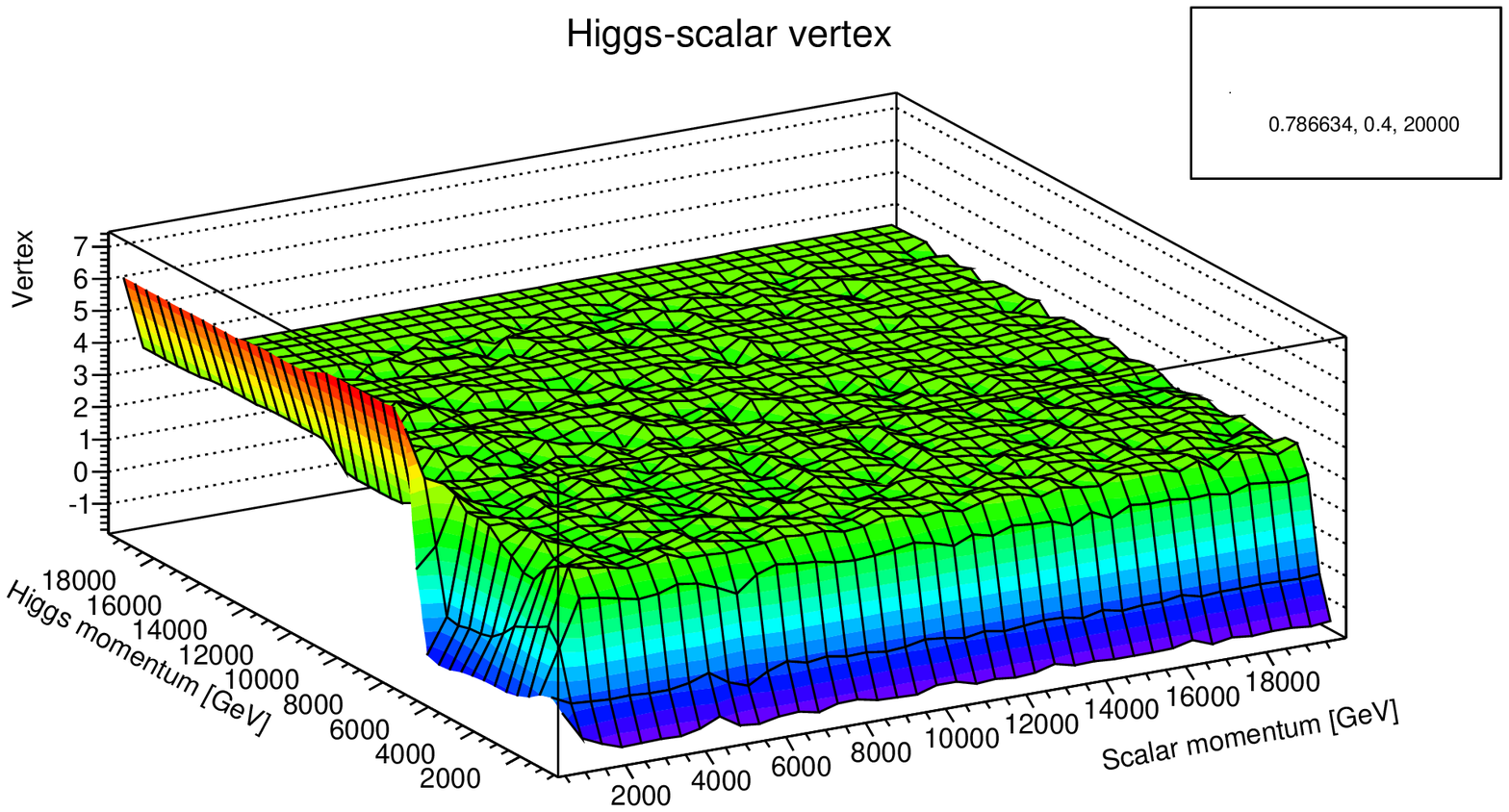}
 }
 \centering
 \parbox{0.47\linewidth}
 {
 \includegraphics[width=\linewidth]{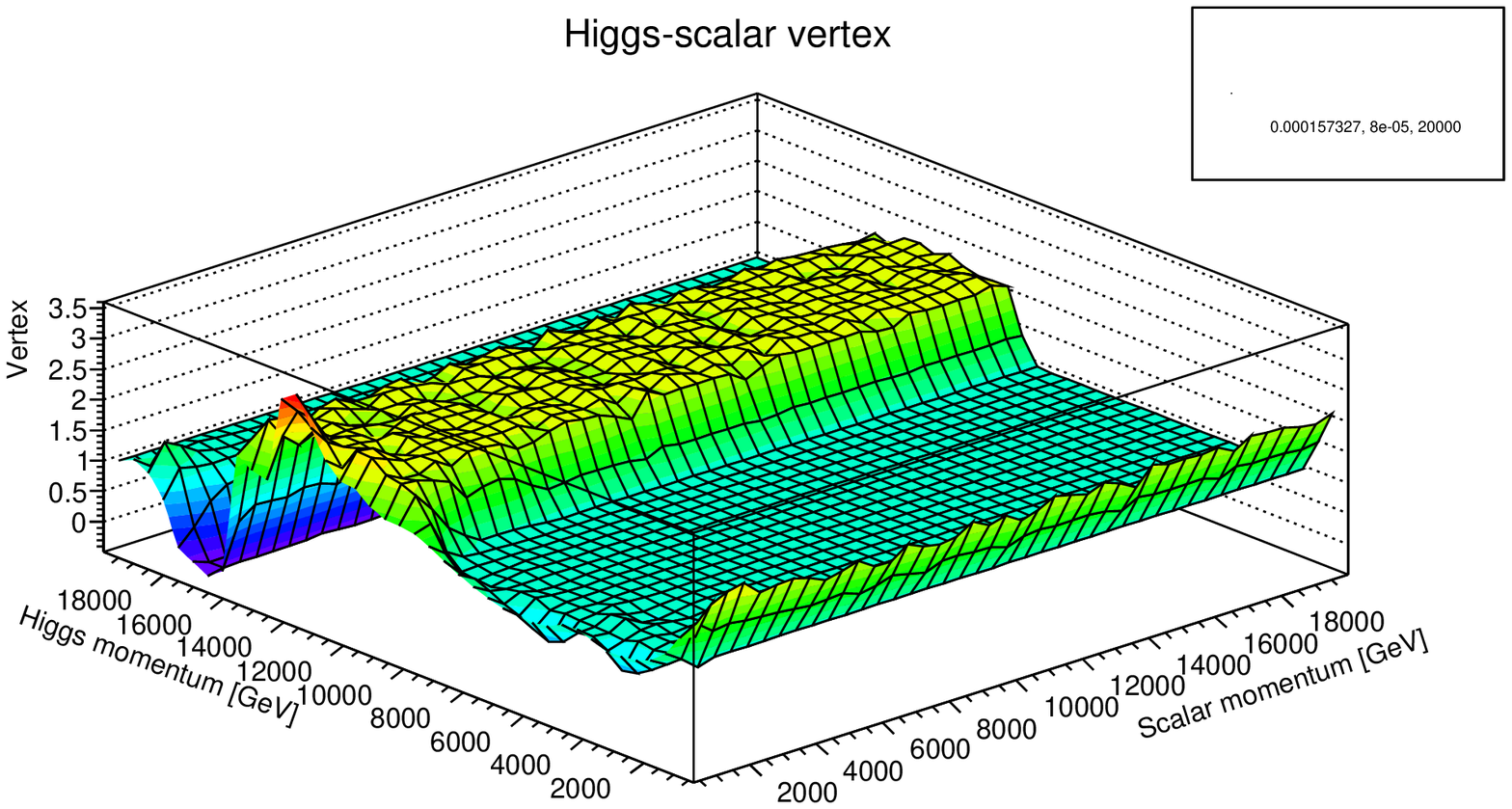}
 }
  \parbox{0.47\linewidth}
 {
 \includegraphics[width=\linewidth]{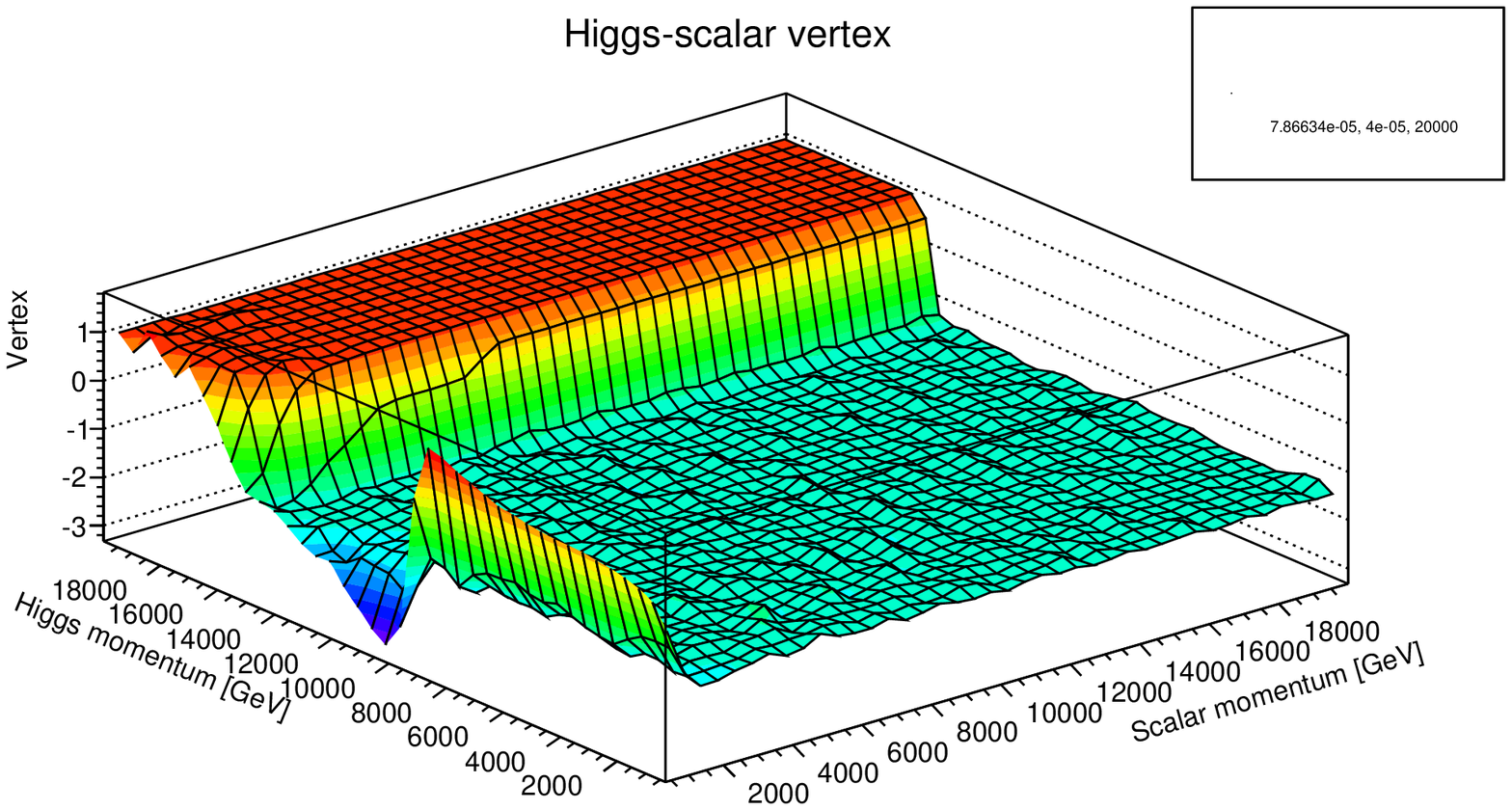}
 }
 \caption{\label{yukl-v-20} Higgs-scalar vertices for different bare couplings are plotted for 20 TeV cutoff value.}
\end{figure}
\begin{figure}
 \centering
 \parbox{0.47\linewidth}
 {
 \includegraphics[width=\linewidth]{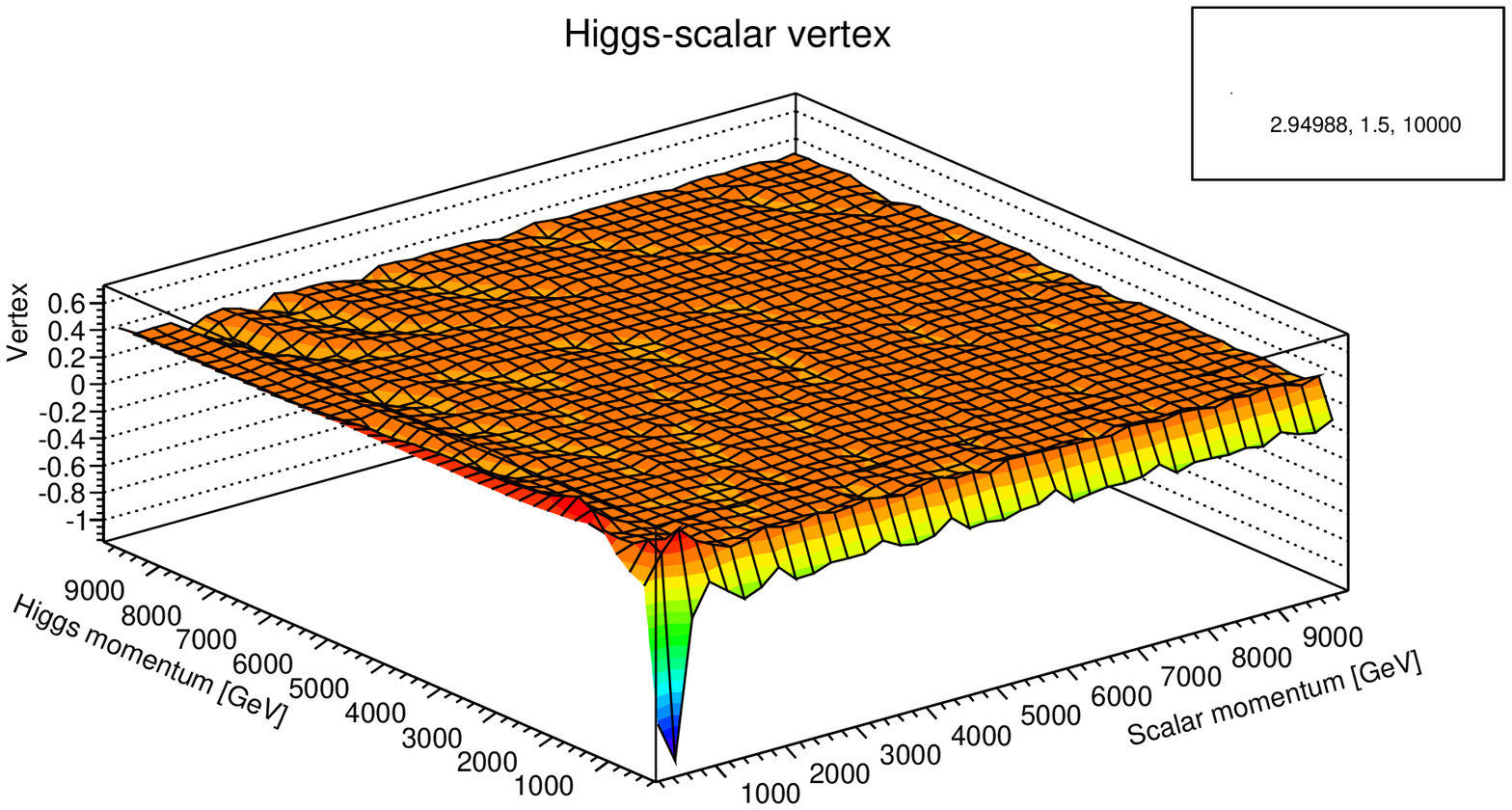}
 }
 \parbox{0.47\linewidth}
 {
 \includegraphics[width=\linewidth]{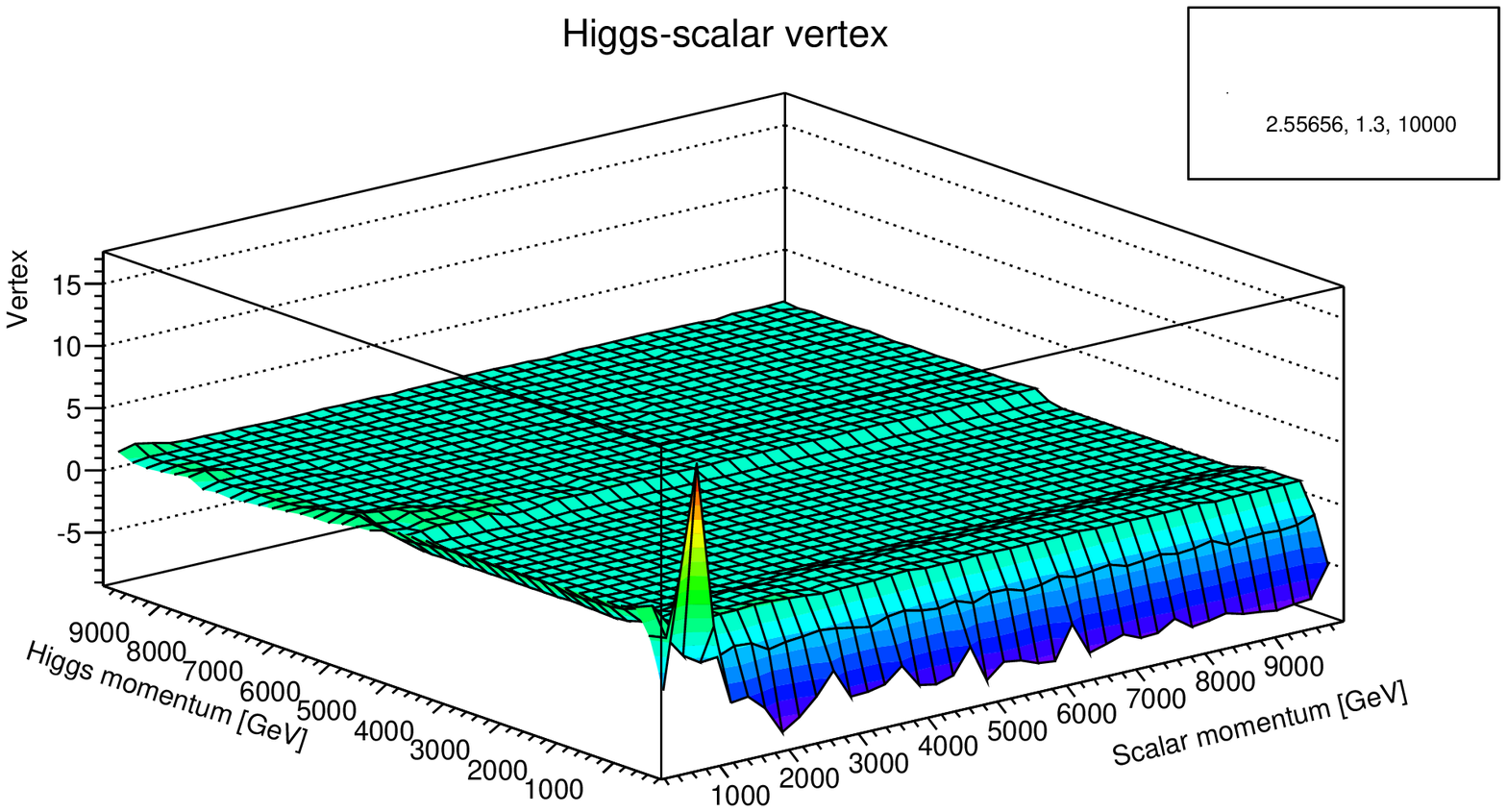}
 }
 \centering
 \parbox{0.47\linewidth}
 {
 \includegraphics[width=\linewidth]{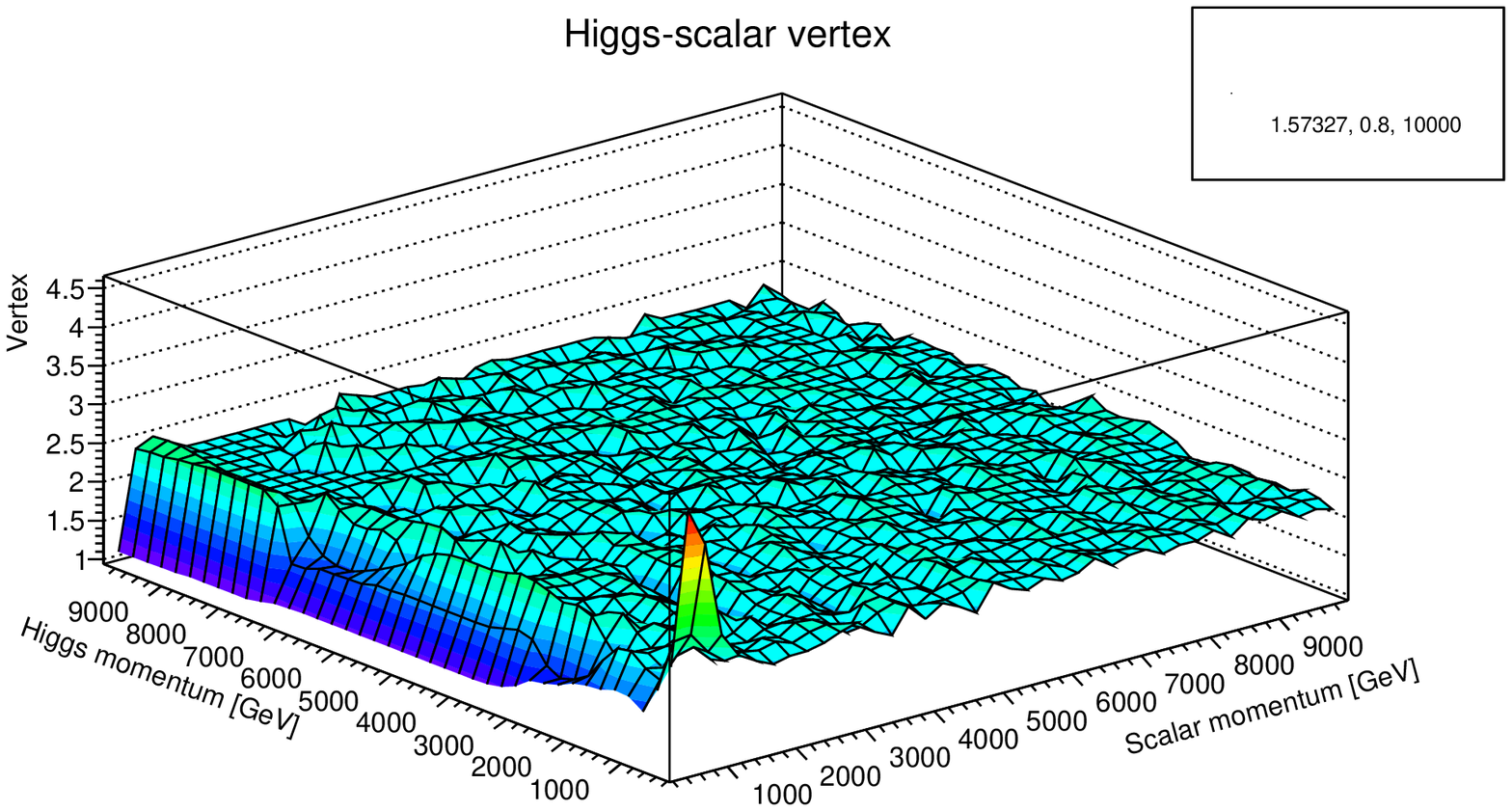}
 }
  \parbox{0.47\linewidth}
 {
 \includegraphics[width=\linewidth]{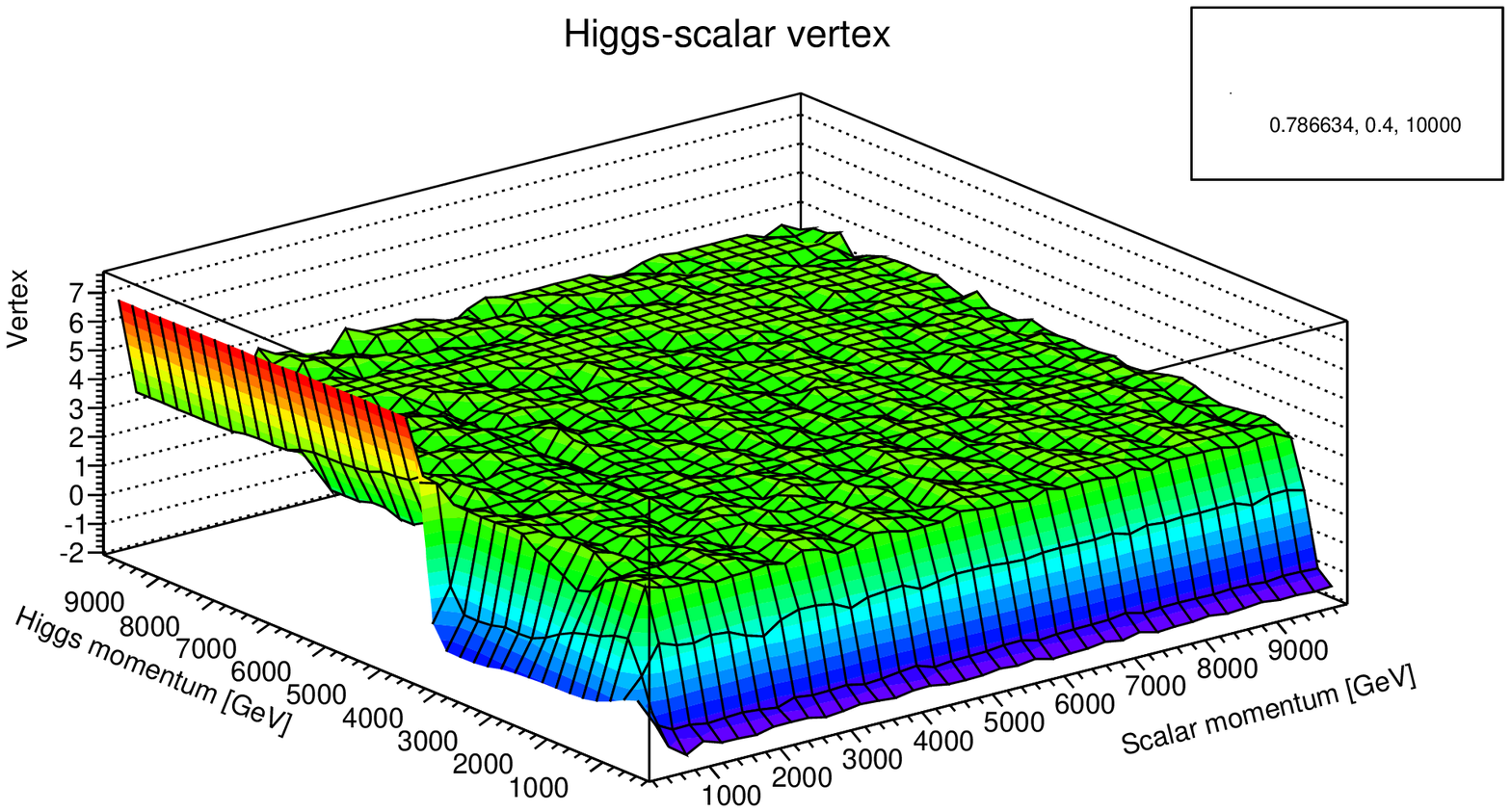}
 }
 \centering
 \parbox{0.47\linewidth}
 {
 \includegraphics[width=\linewidth]{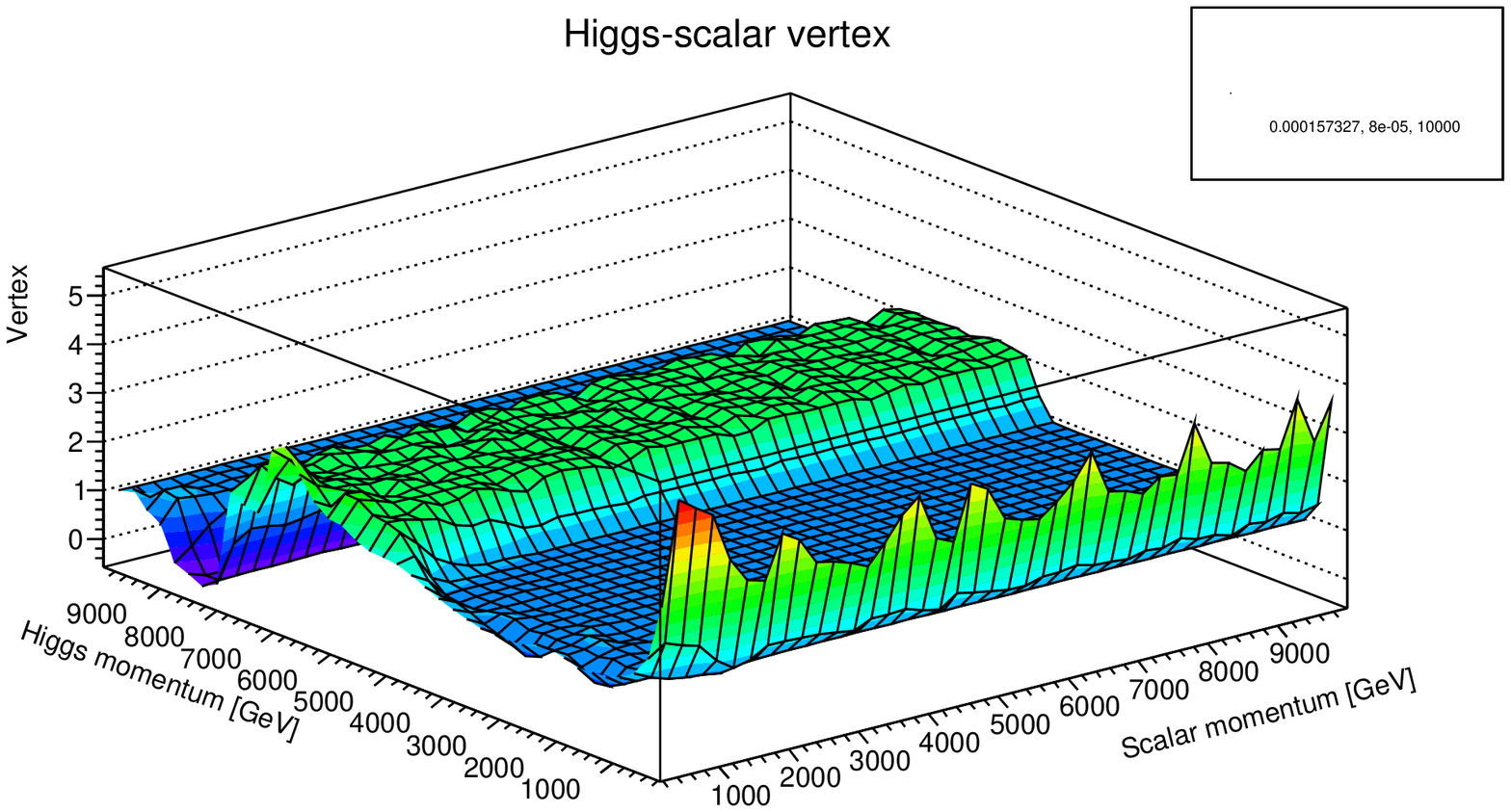}
 }
  \parbox{0.47\linewidth}
 {
 \includegraphics[width=\linewidth]{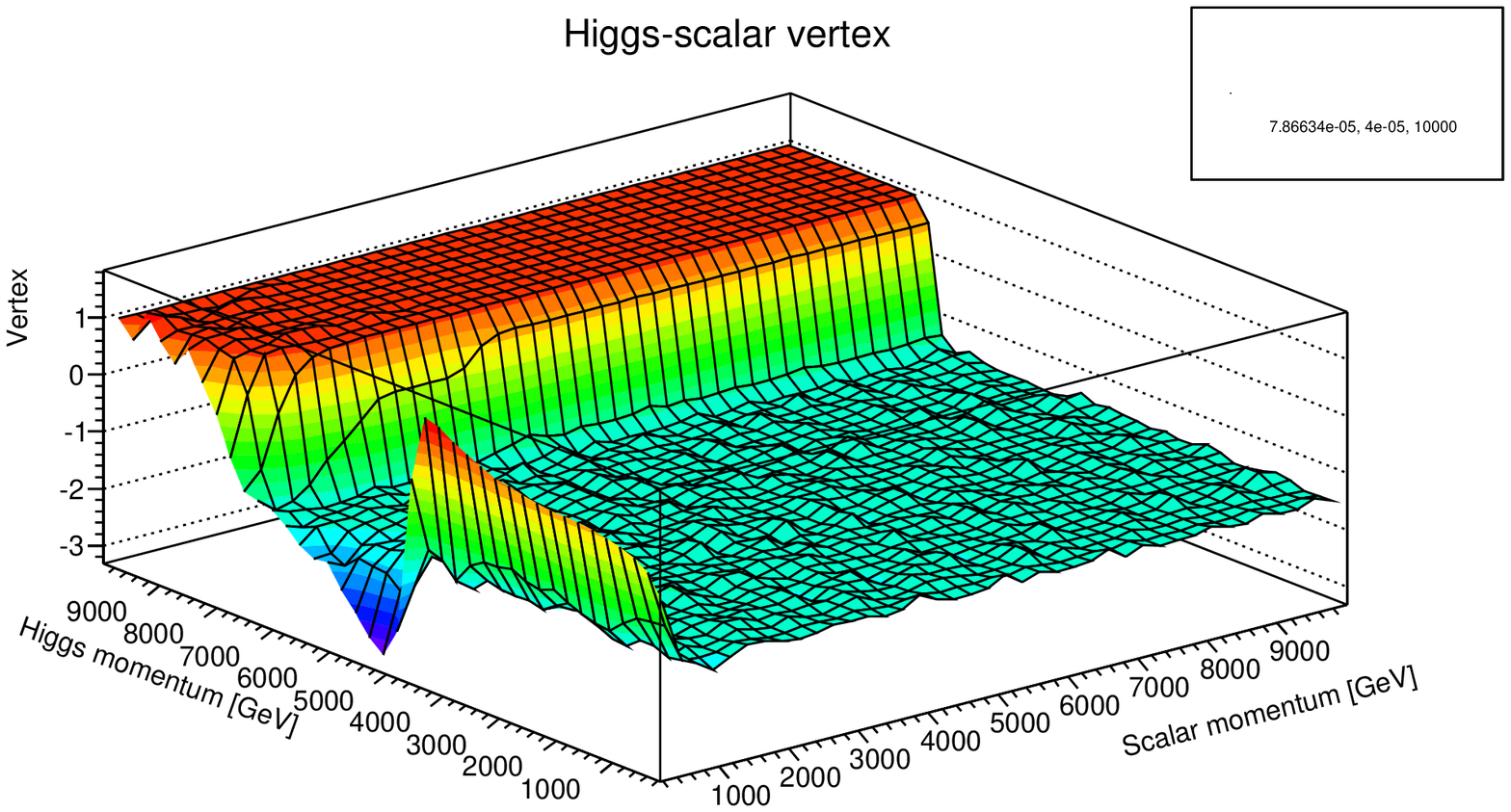}
 }
 \caption{\label{yukl-v-10} Higgs-scalar vertices for different bare couplings are plotted for 10 TeV cutoff value.}
\end{figure}
Three point interaction vertex is the next simplest correlation function after the field propagators. Throughout the computations, as they do not have a DSE for them, they were updated under the constraints of renormalization conditions for the propagators with the momentum configuration mentioned before. The vertex is defined as
\begin{equation} \label{yukl-vtx-def}
\Gamma^{kl}(q,p-q,-p)= \Gamma^{kl}_{(tree)}(q,p-q,-p) V(q,-p)
\end{equation} \label{yukl-vtx-def}
where V(q,p) is the dressing function of the tree level vertex $\Gamma^{kl}_{(tree)}(q,p-q,-p)$ with q and p being the momenta of Higgs and scalar fields, respectively, while $p-q$ is the momentum of the other Higgs (bar) field.
\par
The vertices for the explored region of parameter space here are found to have, more or less, similar features compared to the ones in the previous studies \cite{Mufti:2018xqq}, though there exist slightly more diversity in this region of parameter space. They have both the two plateau feature as well as the flat function behavior over a large spectrum of momentum values. However, transitory behavior is found more often than in previous exploration, see figures \ref{yukl-v-10} - \ref{yukl-v-30} \footnote{As non-perturbative vertices differ from their dressing functions by (constant) tree level expression for vertices, instead of non-perturbative vertices dressing functions are shown in the figures and are referred as vertices.}.
\par
An immediate observation is that, contrary to what was observed before \cite{Mufti:2018xqq}, the vertices no longer converge to unity in the region of high momentum values for most of the coupling values. Hence, even if the perturbative scenario establishes there is no guarantee that it will be with the vertices converging to unity for the region of higher momentum values of both fields. Deviation from perturbative scenario is already not surprising as it is already seen that the corrections to bare scalar mass values coming from scalar self energy terms are significantly higher than the bare values, see figure \ref{masses1} - \ref{masses2}, while Higgs physical mass is found to be the same within $2 \sigma$ (for most of the cases), hence a perturbative picture is not necessarily supported.
\par
As mentioned above, there are many parameters for which scalar squared physical mass receives far more contributions from scalar self energy terms than the values of scalar squared bare mass. As in the previous study, such vertices generally have steep momentum dependence with vertices becoming suppressed or larger in value near the infrared end while retaining insignificant dependence over other values of scalar field's momentum. Despite their dependence over cutoff values, it is observed that the vertices tend to have similar qualitative behavior, see (middle) figures \ref{yukl-v-10} - \ref{yukl-v-30} for an example. Furthermore, there also exist the same steep ascending or descending behavior for Higgs' case. However, this behavior manifests over extremely close to the infrared end which indicates that even if Higgs mass also receives negative contributions from Higgs self energy terms, the effect may not be drastic, see (top) figures \ref{yukl-v-10} - \ref{yukl-v-30} for an example.
\par
For the case of very small coupling values, the vertices tend to strive for maintaining the two plateau behavior. It is also observed that usually the higher plateau is in the higher momentum region, though there also exist transitory vertices as exceptions, see (bottom) figures \ref{yukl-v-10} - \ref{yukl-v-30} for example. Such less abundance of these two plateau vertices are understandable since there are relatively many more cases with squared scalar physical masses with high negative values, hence resulting in a different kind of vertices as mentioned above.
\par
Overall, the interaction vertices are found to be significantly different from their tree level structure in terms of their momentum dependence. Hence, a perturbative treatment may not be the correct tool for investigation in the considered region of parameter space, which was also the deduction in previous studies of the model \cite{Mufti:2018xqq}. Hence, it may be the case that the model does not contain any region in its parameter space suitable for perturbative treatment of the theory.
\section{Conclusion}
In this paper, a variant of Wick Cutkosky model was examined in the parameter space suggested by the ground state of the system with the custodial symmetry intact in the Lagrangian. Despite its simplicity relative to richer QFTs, such as the SM, the theory is found to have important implications.
\par
Recalling that a very different region in the parameter space was examined, it appears clear that in Yuakwa interactions Higgs propagators remain practically unaffected with Higgs physical mass receiving very small contributions beyond the bare mass of, or around, $125.09$. It indicates that, at least strictly in scalar interactions, perturbative picture is highly favored in Higgs interactions in most of the parameter space.
\par
Despite that Higgs is also a (complex doublet) scalar field, singlet scalar is found with entirely different behavior. Its physical mass is found to be strongly depending upon what region in the parameter space is considered. It is known that heavier masses receive relatively smaller contributions beyond the bare mass values. However, for smaller scalar singlet masses, the contributions beyond bare masses are found to be higher by upto 2 orders of magnitude which indicates presence of strong non-perturbative effects which was not observed to this level of severity in previously explored region of parameter space. It leaves the question open, if other interaction terms involving scalar singlet also posses the same features.
\par
The model is found to be generating physical masses which have similar qualitative dependence on bare coupling values as in the ground state of the theory, over a considerable range with higher cutoff values. Given that 3 cutoff values are examined, at this point it is deduced that for higher cutoff values the resemblance may persist in the model. However, the results indicate that the squared physical scalar masses may be produced only at unusually high couplings.
\par
Overall, the vertices retain similar qualitative features as previously found. Hence, the classification set for the vertices in previous study as well as here tend to imply existence of two distinct features of the theory. It does support speculations of 2 phases in the theory. However, distinct behavior of scalar propagators indicate that scalar singlet field in the interactions tends to play vital role than the Higgs field in the Yukawa interaction.
\par
No sign of triviality is found in the parameter space, which further supports the findings in previous exploration. It raises speculations if there exist any islands (regions) in the parameter space which posses such a feature.
\par
As scalar singlet is found to play vital role, particularly in the region of low masses, it begs to explore the model from the perspective of dynamic masses in the theory.
\section{Acknowledgement}
I am extremely thankful to Dr. Shabbar Raza for a number of important advices and valuable discussions throughout the endeavor. This work is primarily supported by faculty research grant from Habib University Karachi Pakistan.
\bibliographystyle{plain}
\bibliography{bib}
\end{document}